\documentclass[review]{elsarticle}

\usepackage[utf8]{inputenc}
\usepackage[a-1b]{pdfx}   
\usepackage{graphicx}
\usepackage{subcaption}
\usepackage{times}
\usepackage{epsfig}
\usepackage{color}
\usepackage{xcolor}
\usepackage{graphicx}
\usepackage{wrapfig}
\usepackage{amssymb, amsmath}
\usepackage{multirow}
\usepackage{framed}
\usepackage{url}
\usepackage{listings}
\usepackage{booktabs} 
\usepackage{enumitem}

\usepackage{hyperref}
\hypersetup{colorlinks=true}

\usepackage[noend]{algpseudocode}
\usepackage{algorithm}
\usepackage{pifont}

\newcommand{\bluecheck}{{\color{blue}\ding{52}}}
\newcommand{\redx}{{\color{red}\ding{54}}}

\journal{Journal of Systems and Software}

\begin{document}

\begin{frontmatter}

\title{Mapping the Structure and Evolution of Software Testing Research Over the Past Three Decades}

\author[ali]{Alireza Salahirad}
\address[ali]{Department of Computer Science \& Engineering, University of South Carolina, USA}
\ead{alireza@email.sc.edu}

\author[greg]{Gregory Gay\corref{mycorrespondingauthor}}
\address[greg]{Department of Computer Science and Engineering, Chalmers $\vert$ University of Gothenburg, Sweden}
\cortext[mycorrespondingauthor]{Corresponding author}
\ead{greg@greggay.com}

\author[ehsan]{Ehsan Mohammadi}
\address[ehsan]{School of Information Science, University of South Carolina, USA}
\ead{ehsan2@sc.edu}

\begin{abstract}  

\noindent\textbf{Background:} The field of software testing is growing and rapidly-evolving.

\noindent\textbf{Aims:} Based on keywords assigned to publications, we seek to identify predominant research topics and understand how they are connected and have evolved. 

\noindent\textbf{Method:} We apply co-word analysis to map the topology of testing research as a network where author-assigned keywords are connected by edges indicating co-occurrence in publications. Keywords are clustered based on edge density and frequency of connection. We examine the most popular keywords, summarize clusters into high-level research topics, examine how topics connect, and examine how the field is changing. 

\noindent\textbf{Results:} Testing research can be divided into 16 high-level topics and 18 subtopics. Creation guidance, automated test generation, evolution and maintenance, and test oracles have particularly strong connections to other topics, highlighting their multidisciplinary nature. Emerging keywords relate to web and mobile apps, machine learning, energy consumption, automated program repair and test generation, while emerging connections have formed between web apps, test oracles, and machine learning with many topics. Random and requirements-based testing show potential decline.

\noindent\textbf{Conclusions:} Our observations, advice, and map data offer a deeper understanding of the field and inspiration regarding challenges and connections to explore.
\end{abstract}


\end{frontmatter}

\section{Introduction}\label{sec:intro}

Software testing refers to the application of input to a system to identify issues affecting its correctness or its ability to deliver services~\cite{Pezze06:testing}. While many quality assurance techniques exist, testing remains the primary means of assessing software quality.

From nearly the beginning of software development as a discipline, researchers and practitioners have reasoned about testing and quality assurance~\cite{Turing49:Checking}. Today, testing is one of the largest areas of software engineering research~\cite{Orso14:STR}, and the field is rapidly evolving as new software and hardware advances are introduced. It is useful, therefore, to understand (a) \textit{what the predominant research topics are of the field}, (b) \textit{how those topics are connected}, and (c), \textit{how the predominant topics have evolved over time}.


``Science of science'' describes a research methodology where text, author, and publication metadata are analyzed using quantitative bibliometric and scientometric techniques~\cite{Fortunato18:SoS,Borgman02:Biblio}. Computational methods, such as text mining and citation analysis, map the topical structure of a research field, enabling the discovery of invisible patterns and relationships in the publications that form that field~\cite{Moed06:Citation,Ding13:References}.

We have applied co-word analysis to visualize and analyze the topology of 35 years of software testing research, based on the author-assigned keywords of Scopus-indexed publications. Co-word analysis yields an undirected network where the nodes---author-assigned keywords---represent targeted research concepts. Weighted edges connect keywords, based on their co-occurrence on publications. Finally, keywords are grouped into clusters, representing densely-connected regions of the network. 

Our analysis maps keywords into dense clusters, from which emerge high-level research topics---themes that characterize each cluster---and makes clear the connections between keywords and topics within and across clusters. It also characterizes the periods in which low-level keywords and high-level topics have emerged---identifying emerging research areas, as well as those where research interest has decreased. The goal of this study is to provide both current and future researchers with perspectives about testing field, built on a quantitative base. For researchers, a snapshot of important disciplinary trends can provide valuable insight into the state of the field, suggest topics to explore, and identify connections (or lack thereof) between keywords and topics that may reveal new insights.  
Among others, we have made the following observations:
\begin{itemize}
    \item Both the most common author-assigned keywords and the keywords that attract the most citations, on average, tend to relate to automation, test creation and assessment guidance, assessment of system quality, and cyber-physical systems.  
    \item These keywords can be clustered into 16 topics: automated test generation, creation guidance, evolution and maintenance, machine learning and predictive modeling, model-based testing, GUI testing, processes and risk, random testing, reliability, requirements, system testing, test automation, test case types, test oracles, verification and program analysis, and web application testing. Below these lie 18 more subtopics.
    \item Creation guidance, automated test generation, evolution and maintenance, and test oracles are particularly multidisciplinary topics, with dense connections to many other topics. Twenty keywords connect topics, reflecting multidisciplinary concepts, common test activities, and test creation information.
    \item Emerging research particularly relates to web and mobile applications, ML and AI---including autonomous vehicles---energy consumption, automated program repair, or fuzzing and search-based test generation. Web applications require targeted testing approaches and practices, leading to emerging connections to many topics. Test oracles are also a rapidly-evolving topic with many emerging connections. ML has emerging potential to support automation.
    \item Research related to random and requirements-based testing may be in decline.
\end{itemize}

We believe that these insights---and the rich underlying networks of keywords---can inspire both current and future researchers in the field of software testing. We additionally make our data available so that others may make their own observations or broaden the horizons of their own research.\footnote{A package containing our data is available at \url{https://doi.org/10.5281/zenodo.7091926}.}

The remainder of this publication is structured as follows. In Section~\ref{sec:bg}, we discuss background concepts and related work. In Section~\ref{sec:method}, we explain our methodology. Section~\ref{sec:results} answers our research questions.  In Section~\ref{sec:advice}, we provide advice on the use of this data, as well as exploratory analyses related to under-explored and missing connections. Section~\ref{sec:threats} details threats to validity. In Section~\ref{sec:conclu}, we offer our conclusions.

\section{Background and Related Work}\label{sec:bg}

\subsection{Bibliometrics and Co-Word Analysis}

Bibliometric analysis is ``the application of mathematical and statistical methods to books and other means of communication''~\cite{pritchard1969statistical}. Bibliometric studies perform quantitative analysis of publications and associated metadata---e.g., keywords, authors, institutions, and citations---to identify themes and patterns within a research field~\cite{de2009bibliometrics}. Such analysis is often combined with mapping techniques to visualize hidden structures in the metadata of a particular field~\cite{donthu2020forty}. The most common analysis methods used include citation-based, co-word (also known as keyword co-occurrence), and co-authorship analysis~\cite{van2014visualizing}. We focus on co-word analysis. 

In co-word analysis, natural language processing and text mining techniques are used to discover the most meaningful noun phrases in a collection of documents and visualize their meaning in a two-dimensional map~\cite{Peters93:CoWord}. In this map, co-occurring terms are connected, with ``closer'' placement resulting from stronger co-occurrence. Co-word analysis is generally based on the number of research publications where two keywords are used together to describe the research performed~\cite{whittaker1989history}. Because keywords succinctly capture the context of a publication, co-word analysis is an effective method of revealing connections between publications~\cite{su2010mapping} and identifying trends in a field~\cite{Ding13:References}.

Scholars have previously used co-word analysis to depict the structure of fields including renewable energy~\cite{Romo13:co}, global warming~\cite{Werner17:CoWord}, nanoscience and nanotechnology~\cite{Mohammadi12:CoWord}, human computer interaction~\cite{Liu14:CHICoWord}, and big data~\cite{Liao18:MedicalCoWord,Mohammadi20:BigData}. Our study is the first to apply such techniques to software testing.

\begin{table}[!t]
\resizebox{\columnwidth}{!}{
\begin{tabular}{lcccccccccc}
\hline
\textbf{Topic} & This Study & \cite{Garousi16:SEBib} & \cite{Garousi17:Quantity} & \cite{Karanatsiou19:SE,Wong08:Scholars,Wong09:Scholars,Wong11:Scholars} & \cite{Garousi10:Canada,Garousi15:SETurkish} & \cite{Farhoodi13:SciSoft} & \cite{deFreitas11:SBSE} & \cite{Harrold00:roadmap} & \cite{Bertolino07:Achievements} & \cite{Orso14:STR} \\  \hline
Field & Testing & All SE & All SE& All SE & All SE & Sci. SW & SBSE & Testing & Testing & Testing\\ \hline 
Method & Quan., Qual. & Quan. & Quan. & Quan. & Quan. & Quan., Qual. & Quan. & Qual. & Qual. & Qual. \\ \hline 
Research Topics & \bluecheck & \bluecheck & \redx & \redx & \redx & \bluecheck & \redx & \redx & \bluecheck & \bluecheck\\ \hline
Topic Connections & \bluecheck & \redx & \redx & \redx & \redx & \redx & \redx & \redx & \redx & \redx\\ \hline
Keyword Clustering & \bluecheck & \redx & \redx & \redx & \redx & \redx & \redx & \redx & \redx & \redx\\ \hline
Keyword Connections & \bluecheck & \redx & \redx & \redx & \redx & \redx & \redx & \redx & \redx & \redx\\ \hline
Popular Keywords & \bluecheck & \redx & \redx & \redx & \redx & \redx & \redx & \redx & \redx & \redx\\ \hline
Emerging Topics & \bluecheck & \redx & \redx & \redx & \redx & \redx & \redx & \bluecheck & \bluecheck & \bluecheck \\ \hline
Declining Topics & \bluecheck & \redx & \redx & \redx & \redx & \redx & \redx & \redx & \redx & \redx \\ \hline
Underexplored Con.s & \bluecheck & \redx & \redx & \redx & \redx & \redx & \redx & \redx & \redx & \redx \\ \hline
Potential Connections & \bluecheck & \redx & \redx & \redx & \redx & \redx & \redx & \redx & \redx & \redx \\ \hline
Popular Papers & \redx & \redx & \redx & \redx & \redx & \redx & \bluecheck & \redx & \redx & \redx\\ \hline
Top Authors & \redx & \redx & \redx & \bluecheck & \bluecheck  & \bluecheck & \bluecheck & \redx & \redx & \redx\\ \hline
Author Location & \redx & \bluecheck & \bluecheck & \bluecheck & \bluecheck  & \bluecheck & \redx & \redx & \redx & \redx\\ \hline
Pub. Venue & \redx & \redx & \bluecheck & \redx & \redx  & \bluecheck & \bluecheck & \bluecheck & \redx & \redx\\ \hline
\end{tabular}}
\caption{
Comparison of our study to other related work, based on the research field, methodologies, and analyses performed. 
} 
\label{tab:related_work} 
\end{table}

\subsection{Bibliometrics and Software Engineering}

Our study is the first to apply scientometric or bibliometric techniques to the software testing field. However, bibliometric techniques have been applied to other aspects of software engineering (SE). In Table~\ref{tab:related_work}, we contrast our study to related work. Below, we further elaborate on the specific studies. In general, we focus on analysis of research topics and the connections between topics, and do not analyze authorship trends. Our focus and chosen analysis methods enable a deep characterization of the connections between topics and low-level publication keywords in software testing.

Garousi and M\"{a}ntyl\"{a} performed a bibliometric analysis of more than 70,000 general SE publications, finding that the most popular research topics were web applications, mobile and cloud computing, industrial case studies, source code, and automated test generation~\cite{Garousi16:SEBib}. Our identified research topics include all of these except source code---which is subsumed by other topics---and case studies. In our study, case studies would be categorized based on the problems they address. They also found that a small number of large countries produce the majority of publications, while small European countries are proportionally the most active in the field. 

Garousi and Fernandes used the same set of publications to assess questions related to quantity versus the impact of SE research~\cite{Garousi17:Quantity}. They broadly found that journal articles have more impact than conference publications and that publications from English-speaking researchers have more visibility and impact. Both studies also used Scopus to gather publications, but had a different focus from our study (all of ``software'', rather than software testing). The studies also differ in their analysis methods. Rather than co-word analysis, the authors of both studies used citation-based analyses. Co-word analysis allows examination of the connections between topics.  

Karanatsiou et al. targeted SE publications from 2010-2017 for analysis, identifying top institutions and scholars from this period~\cite{Karanatsiou19:SE}. Wong et al. did the same for the periods of 2001-2005~\cite{Wong08:Scholars}, 2002-2006~\cite{Wong09:Scholars}, and 2003-2007 and 2004-2008~\cite{Wong11:Scholars}. Garousi et al. also performed bibliometric analysis, specifically, on the SE research communities in Canada~\cite{Garousi10:Canada} and Turkey~\cite{Garousi15:SETurkish}. These studies differ from our own in their focus on the authors of publications, rather than research topics.  

Farhoodi et al. reviewed literature related to scientific software, finding that many SE techniques are being applied in the field and that there is still a need to explore the usefulness of specific techniques in this context~\cite{Farhoodi13:SciSoft}. Their focus differs in both the analysis techniques, and in their focus on a specific software domain. In Section~\ref{sec:topics_emerging}, we do observe the emergence of testing research related to scientific software. 

De Freitas and de Souza performed a bibliometric analysis on the first ten years of research in search-based software engineering---the use of optimization techniques to automate tasks~\cite{deFreitas11:SBSE}. They identified the most cited papers, most prolific authors, and analyzed the distribution of the SBSE publications among conference proceedings, journals, and books. They described networks of collaborations and distributions of publications in various venues and identified the distribution of the number of works published by authors. Their study differs from ours in its focus on a particular research domain, as well as its focus on authors and venues over research topics. 


\subsection{Other Related Work}

Purely qualitative analyses of testing research have also been performed. In Table~\ref{tab:related_work}, we contrast our study to those discussed below. None of these studies perform a full summarization or mapping of the testing field. Instead, they point out research areas that are emerging or that have had a major impact. The topics they discuss tend to form a subset of those in our characterization of the field. In addition, our quantitative analysis methods enable elaborate analyses of the field and the connections between topics not explored in these studies. 

Harrold, in 2000, examined past research to identify areas of focus for future research~\cite{Harrold00:roadmap}. These areas include improvements in integration testing, use of pre-code artifacts (e.g., specifications) to plan and implement testing activities, development of tools for estimating, predicting, and performing testing on evolving systems, and process improvements. Many of these predictions are now established topics in our map, such as black box testing, evolution and maintenance, and processes and risk.

Bertolino provided a summary of testing research in 2007, and identified achievements in the testing process, reliability testing, protocol testing, test criteria, object-oriented testing, and component-based testing as major advances~\cite{Bertolino07:Achievements}. She identified outstanding challenges related to testing education, testing patterns, cost of testing, controlled evolution, leveraging users, test input and oracle generation, model-based testing, and testing of specialized domains, among others. Many of her achievements and challenges appear in our map as either keywords or full research topics. 

Orso and Rothermel assessed research performed in the field between 2000-2014, asking colleagues what they believed were the most significant contributions and the greatest challenges and opportunities~\cite{Orso14:STR}. The research contributions were categorized into the areas of automated test generation, testing strategies, regression testing, and support for empirical publications. The first three of those areas reflect research topics in our map. Challenges identified included better testing of modern, real-world systems, generation of test oracles, analysis of probabilistic programs, testing non-functional properties (e.g., performance), testing of specialized domains (e.g., mobile), and leveraging of the cloud and crowd. Some of these challenges---e.g., mobile and performance testing---are now research topics in our map.

\section{Methodology}\label{sec:method}

Software testing is one of the most popular and fast-growing areas of software engineering research~\cite{Orso14:STR}. Although there are many surveys, mapping studies, and systematic literature reviews on individual topics, there is a lack of quantitative examination of the field as a whole---mapping research topics and their connections. 

Our primary goal is to provide and analyze a ``map'' of the field of software testing, based on the many distinct research keywords that form the field and the connections between these keywords, linked through research publications. Our mapping is based on a quantitative method, co-word analysis, that places co-occurring phrases---in our case, author-supplied keywords---in a network. Within this network, keywords appear as nodes, with weighted edges indicating how often keywords are linked in publications. Sets of strongly co-occurring keywords form distinct clusters. This network structure offers a quantitative method to characterize the research field, which can be used as the basis of both qualitative and quantitative analyses.  

Using this map, we examine how keywords are linked into \textit{clusters}, characterize clusters using high-level research \textit{topics}, examine the connections between keywords \textit{within} and \textit{across} clusters, and examine how interest in particular keywords and topics have changed \textit{over time}. 
Specifically, we address the following research questions:
\begin{itemize}
    \item[\textbf{RQ1:}] What are the most popular individual keywords in software testing, as indicated by the number of publications or citations?
    \item[\textbf{RQ2:}] What topics characterize the keywords connected \textit{within} each cluster in the map?
    \item[\textbf{RQ3:}] How are keywords and research topics most strongly linked \textit{across} clusters?
    \item[\textbf{RQ4:}] What keywords, topics, and connections have emerged or grown in popularity over the past five years?
    \item[\textbf{RQ5:}] Which keywords and topics have shown the greatest decline in interest? 
\end{itemize}

We begin, in \textbf{RQ1}, by examining the individual keywords targeted by authors. We are interested in identifying which keywords have been selected most often, and which receive the most citations per publication on average. We then move into analyses and characterization of the \textit{connections} between keywords. 

The goal of \textbf{RQ2} is to summarize each cluster. Keywords \textit{within} a single cluster are highly interconnected, providing a basis for identifying \textit{research topics} that encapsulate connected keywords. A topic as a keyword or phrase that connects multiple keywords. For example, ``automated test generation'' is not just a single keyword, but also a topic that connects other keywords such as ``ant colony optimization'' and ``genetic algorithm'' within the same cluster.\footnote{Both are algorithms often used to generate tests, linking all three keywords as part of the same topic.} \textbf{RQ3}, then, focuses on the connections \textit{across} clusters, and characterizes how keywords and research topics connect. 

\textbf{RQ4} and \textbf{RQ5} focus on an additional dimension, the average age of publications associated with each keyword. In \textbf{RQ4}, we identify keywords, topics, and connections between keywords and topics that have emerged or grown in popularity in the past five years. In \textbf{RQ5}, we examine keywords and topics with the oldest average date of publication---those with a potential decline in interest. These emerging and declining concepts offer insight into how the field is evolving. 

To answer these questions, we (1) collected publications related to software testing (Section~\ref{sec:collection}), (2) constructed a map, using co-word analysis, of clusters of connected keywords  (Section~\ref{sec:mapping}), (3) removed unrelated or redundant topics (Section~\ref{sec:cleanup}), and (4), analyzed the map and underlying data (Section~\ref{sec:analysis}).

\subsection{Data Collection}\label{sec:collection}

To gain an inclusive overview of software testing, we gathered publications from the Scopus database. Scopus is a comprehensive meta-database, covering many conference and journal publication venues. We retrieved all publications returned for the search term \textit{``software testing''} on September 26, 2020. Only publications published in English were used. This collection included 57,233 publications. 

\begin{figure}[!t]
    \centering
    \includegraphics[width=\linewidth]{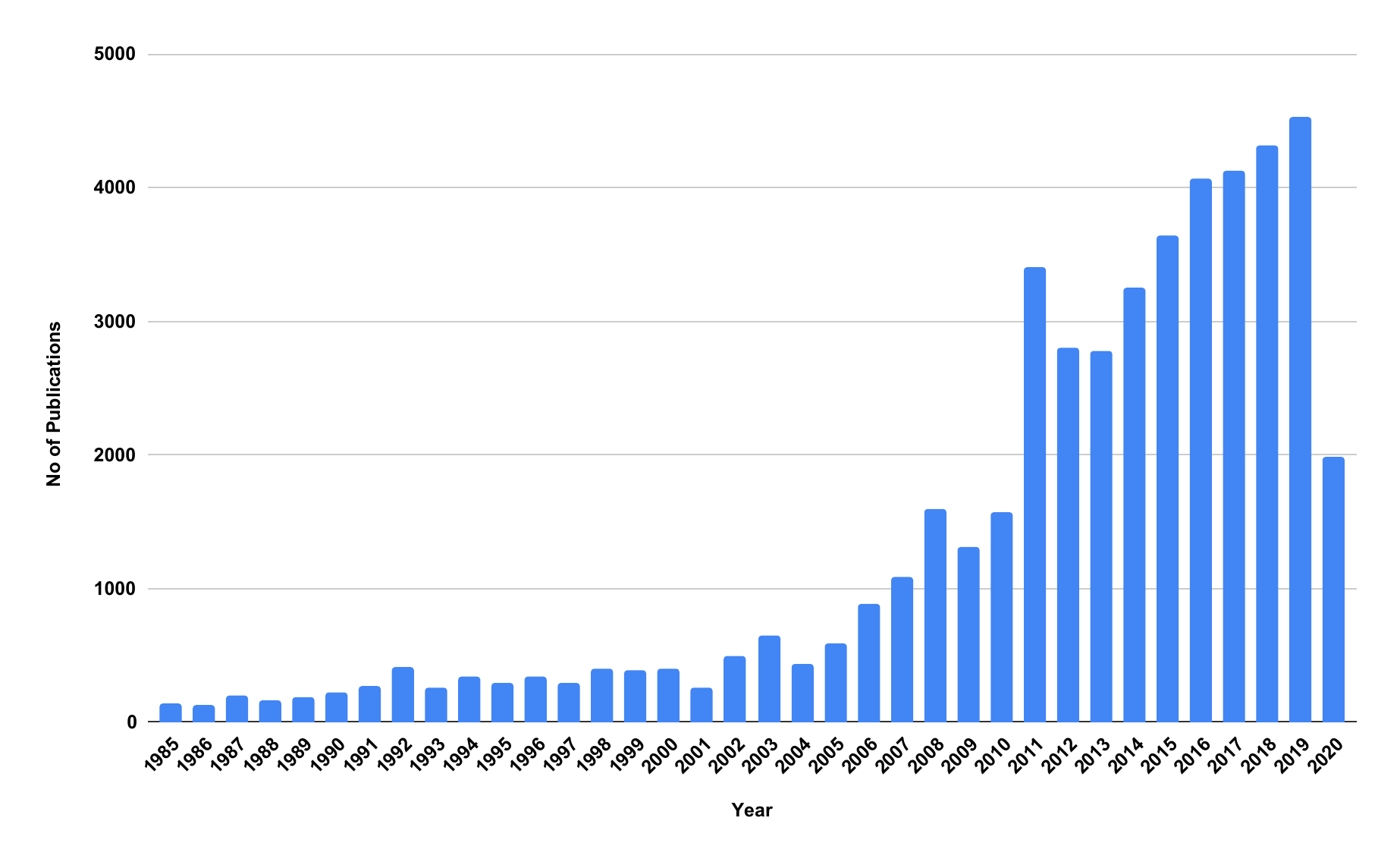} \vspace{-25pt}
    \caption{Number of publications per year retrieved from Scopus.}
    \label{fig:numPapers}
\end{figure}

Following a manual cleaning stage (see Section~\ref{sec:cleanup}), 49,802 publications were included, including 36,774 conference papers, 11,640 journal articles, and 1,388 other articles. Figure~\ref{fig:numPapers} gives an overview of the number of publications published per year. Our aim was to capture a representative sample of the field, not all possible articles on software testing. When we quote specific numbers of publications, these numbers should not be taken as absolutes, but as the approximate commonality of a topic.

For each study, we gathered the title, author data (names, affiliations, locations), keywords, publication date, venue metadata (e.g., publisher, venue, volume, page numbers), number of citations, DOI, link, and language.

\subsection{Map Construction}\label{sec:mapping}

To map testing research, we used co-word analysis~\cite{Peters93:CoWord}. Co-word analysis is a natural language processing method that extracts important phrases from a textual dataset and identifies their relationships in a network based on the number of times that two terms co-occur together in all documents. This technique assumes that terms that co-occur more often are more strongly related to each other. As a result, all identified terms are classified into clusters using co-occurrence to measure term similarity and depict the extracted terms and their relationship in a two-dimensional visualization. 

We used VOSviewer (Visualization of Similarities Viewer) to analyze the collected data. VOSviewer is a tool that creates maps based on network data~\cite{Van10:VOSviewer}. These maps provide visualizations that allow researchers to explore items and relationships. There are various methods for establishing connections between items in these networks, including co-authorship, co-occurrence, citation, bibliographic coupling, and co-citation.

We tested title, abstracts, index keywords, and author-supplied keywords as the unit of analysis and found that author-supplied keywords are the most promising way to identify research topics and their connections. 

In this analysis, we considered 20 as the minimum threshold of keyword occurrences. This threshold places a minimal barrier before a keyword is ``important'' enough to incorporate. Keywords appearing in fewer than 20 publications were omitted. This threshold was chosen after experimentation as a way to control the level of noise and difficulty of interpretation of the dataset and map, while still avoiding potential loss of interesting and emerging topics. We then iteratively removed keywords that were unrelated to software testing (e.g., publications that used software as part of classroom testing) and merged redundant keywords (e.g., ``automated test generation'' and ``automated test case generation'')---see Section~\ref{sec:cleanup}---leaving a final set of 406 keywords. 


VOSviewer produces maps based on a co-occurrence matrix---a two-dimensional matrix where each column and row represents an item---a keyword, in our case--and each cell indicates the number of times two keywords co-occur. This map construction consists of three steps. In the first step, a similarity matrix is created from the co-occurrence matrix. A map is then formed by applying the VOS mapping technique to the similarity matrix. Finally, the map is translated, rotated, and reflected. We provide technical details on VOSviewer's algorithm in~\ref{sec:vosviewer}. 


In VOSviewer, a map is visualized in three ways: The network visualization, the overlay visualization, and the density visualization~\cite{VanEck14:Visualizing}. We have used the network and overlay visualizations in this study, as well as the raw underlying data. 

\begin{figure}[!t]
\centering
\includegraphics[width=0.8\textwidth]{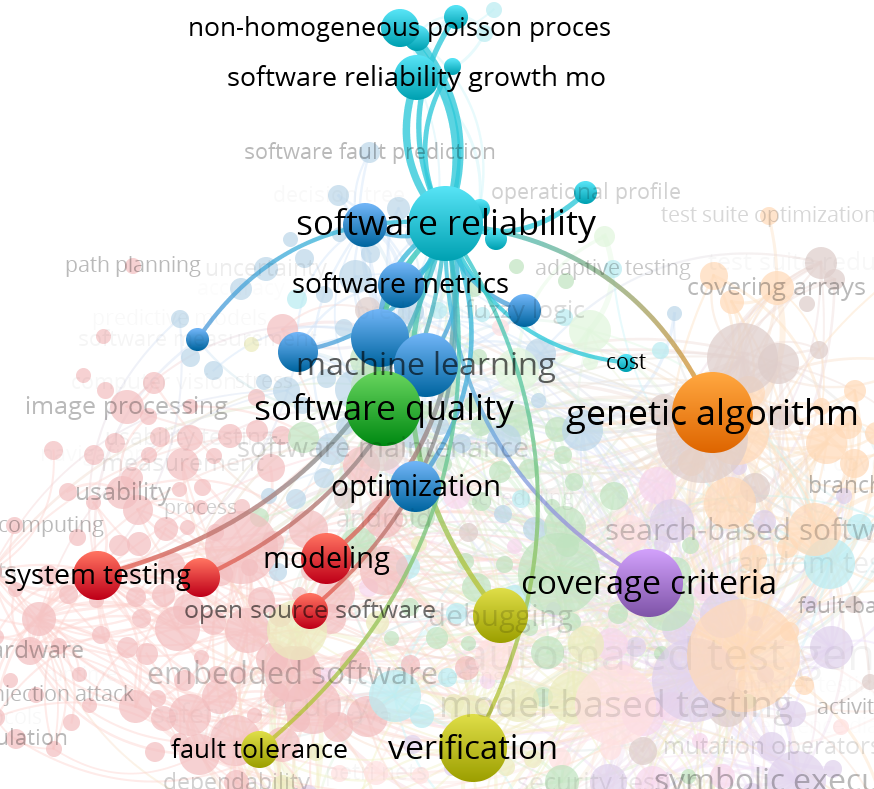}
  \caption{Topics associated with software reliability}
    \label{fig:reliability} 
\end{figure}

The network visualization is the standard view, displaying clusters of related items, connected with edges based on their co-occurrence. Figure~\ref{fig:topics} shows the full network visualization that is produced. In Figure~\ref{fig:reliability}, we highlight a small portion to explain how to interpret the map data.

In this map, each node is a keyword. Figure~\ref{fig:reliability} focuses on the keyword ``software reliability''. All keywords with a sufficiently strong connection to the targeted keyword are highlighted, while unrelated keywords are made partially transparent. The size of a node is based on the number of occurrences of the keyword. In Figure~\ref{fig:reliability}, software reliability is targeted in approximately twice as many publications as ``optimization''. 

Keywords are organized into clusters according to the process described above. Individual keywords can be linked across different clusters. However, the keywords within a cluster tend to be very closely linked with several other keywords within the same cluster. The color of the node indicates its cluster. In Figure~\ref{fig:reliability}, software reliability is marked in light blue, and other nodes with the same color belong to the same cluster (e.g., ``software reliability growth model''). 

Keywords that co-occur in publications are illustrated with an undirected edge. The thickness of the edge indicates how many publications have targeted both keywords. In Figure~\ref{fig:reliability}, software reliability and software reliability growth model share a stronger association (co-occurring in 35 publications) than software reliability with coverage criteria (5 publications). A user-controlled threshold determines the minimum connection strength for visible edges. We used the default, four publications, to control the level of noise when using the visualization for interpretation. When performing quantitative analyses, we consider all connections, regardless of strength.

The overlay visualization uses colors to indicate certain properties of a node, like the average number of citations that publications targeting a keyword have received, instead of using colors to show the cluster. In our case, we use this visualization to analyze the average age of publications targeting a keyword. 

\subsection{Data Cleanup}\label{sec:cleanup}

The initial data included keywords that were either redundant or irrelevant:
\begin{itemize}
    \item There are a small number of keywords unrelated to software testing, as the initial sample was gathered using a broad search string. For example, there were keywords related to software-based student examination or software-based testing of hardware. Additional keywords are either too generic to be considered as specific research concepts---e.g., ``software testing''---or are research-related terms---e.g., ``case study'', ``empirical study''. 
    \item Multiple keywords can refer to the same concept, 
    and can be streamlined into a single keyword---e.g., ``automated test generation'' and ``automated test case generation''. The same keyword can appear in singular and plural form---e.g., ``test case'' and ``test cases''. There are also American and British English spellings (e.g., ``prioritisation'' and ``prioritization''). 
\end{itemize}

To handle irrelevant and redundant keywords, we performed an iterative process. The authors discussed each keyword and came to a consensus. We removed irrelevant keywords from the map, as well as those considered too broad or generic. We removed publications targeting only those keywords, but retained publications that had additional keywords that remained in our set. 

We merged redundant keywords. In performing this process, we limited merging to cases where a redundancy was obvious---primarily pluralization and British/American English. This was to limit the risk of biasing the underlying data that we are using to draw conclusions. We discussed each keyword and its alternatives, and came to a consensus on which keyword to use in all cases. We then replaced the merged keywords with the final keyword for each study and recreated the maps. We performed this process multiple times until we were satisfied that redundant keywords did not remain. 

\subsection{Data Analyses}\label{sec:analysis}

\smallskip\noindent\textbf{RQ1 (Popular Keywords):} To identify the most common keywords, we sort the keywords by the number of publications that targeted that keyword, and examine those that are targeted in $\geq$ 0.50\% of publications, or $\geq$ 250 publications. This threshold was chosen by examining the drop-off in significance over the ten most popular keywords and by considering the trade-off between clarity and giving a thorough impression of the testing field. A total of 20 keywords fall above this threshold (4.93\% of keywords). 

We also have examined which keywords have received the most citations per publication, on average. Here, we examine all keywords with an average number of citations $\geq 20$. This threshold yields 23 keywords, and was chosen because it yields a similar quantity to the number of most common keywords, enabling clearer comparison.

\smallskip\noindent\textbf{RQ2 (Characterization of Clusters into Topics):} We perform a qualitative characterization to summarize the field of software testing, supported by the clusters. We perform this summarization by assigning a small number of high-level ``topics'' to each cluster---keywords or phrases that connect multiple keywords. We have chosen these topics based on our interpretation of the keywords within each cluster. For example, ``performance testing'' is a keyword that connects, e.g., ``load testing'', ``cloud computing'', and ``cloud testing''.\footnote{``load testing'' and ``cloud testing'' are forms of performance testing that target ``cloud computing''.} Because that keyword summarizes many other keywords and connections within that cluster, ``performance testing'' can also serve as a topic that describes the cluster as a whole. 

Some clusters can be summarized by a single topic, while others represent multiple topics. There is no case where \textit{all} keywords are connected to \textit{all} other keywords in the same cluster. Often, two keywords are only indirectly connected through other keywords---e.g., ``random testing'' is connected to  ``reliability'' and ``adaptive random testing'', but the latter two are not directly connected in our sample.

We often observed small sub-groupings of keywords within a topic. In such cases, we assign both a topic and a \textit{subtopic}. For example, the topic of test creation guidance can be broken into four distinct types of creation guidance. A keyword can also belong to multiple topics or subtopics, linking topics within a cluster. 

To assign topics, all authors examined the keywords within a cluster. We then grouped keywords into one or more groupings. To be grouped, keywords must have a direct or indirect (via a shared keyword) edge. This grouping was made based on our experience, literature, and the map data. We grouped keywords if they were used to perform the same activity, were a technology used to perform an activity, were a source of data for an activity, or had some other clear shared purpose. The proposed groupings were discussed until a consensus could be reached. We then identified either a keyword or concept that characterized each grouping. We discussed the options and reached a consensus on the topic to assign. In cases where a grouping could be split into smaller, but still distinct, groupings, subtopics were identified. 

As all keywords must be grouped into a cluster, there are situations where a small number of individual keywords do not relate to the topics assigned to that cluster. We have tried to select topics that are as inclusive as possible. 

\smallskip\noindent\textbf{RQ3 (Connections Across Clusters):} In this question, we are interested in characterizing how keywords and topics are connected \textit{across} clusters. To do so, we have examined two concepts---measurement of the \textit{density of connections between clusters}, and the identification of \textit{keywords that are connected to many other keywords}. 

\noindent\textit{Connection Density:} We can examine how clusters are connected by identifying the cases where the largest percentage of possible connections exist between keywords in two clusters, A and B: 
\begin{equation}
\frac{(\text{Number of Connections Between Keywords in Clusters A and B})}{(\text{Number of Keywords in Cluster A}) * (\text{Number of Keywords in Cluster B})}
\end{equation}
\noindent A measurement of 1.00 means that all keywords in Cluster A are connected to all keywords in Cluster B.\footnote{We employ the density because different clusters contain different numbers of keywords. This measurement offers a fairer basis of comparison than the raw number of connections between pairs of clusters.}

To identify the most densely interconnected clusters, we measured the connection density between all pairs of clusters. We then focus on the connections with a density $\geq 0.12$ (where at least 12\% of all possible connections exist). This threshold was chosen after examination of the measurements---23 of the 45 pairs of clusters ($~50\%$) have a connection density above this threshold.

\noindent\textit{Connecting Keywords:} Some keywords are singular research concepts, while others serve as ``connecting keywords'' that link many keywords together. We further characterize connections across clusters by identifying these connecting keywords. 

To identify these keywords, we measure the number of keywords that each keyword co-occurs with outside of the cluster where that keyword is located. We analyze all keywords connected to $\geq 100$ keywords in external clusters. This threshold was chosen as it yielded the same number of keywords as the threshold for the most popular keyword in RQ1 (20 keywords), enabling direct comparison. 

\smallskip\noindent\textbf{RQ4 (Emerging Keywords, Topics, and Connections):} In this question, we are interested in identifying emerging trends in testing research. We can do this by examining individual keywords, research topics, and connections between keywords.

We have classified any keywords with an average date of publication \textit{later than June 2015} as our set of emerging keywords. This captures an approximate five year period ending with the date we took our sample of publications. A recent date implies one of two things about a keyword: (1) this is a new keyword, or (2), this is a older keyword that has received more attention in recent years.  



We are also interested in examining the connections between keywords and topics, as related to the set of emerging keywords. There are a total of 2,029 connections where at least one of the connected keywords is an emerging keyword, of which 1,412 are cross-cluster connections and 617 are within-cluster connections. To focus our analysis, we focus on the cross-cluster connections, allowing us to also examine and characterize the emerging connections between topics. 

To identify a subset of those 1,412 connections for further exploration, we use the cross-cluster connection density to identify the pairs of clusters with the highest proportion of emerging connections. We have selected the ten pairs of clusters with the highest proportion of emerging connections for further examination, corresponding to a threshold of $\geq 3.5\%$ of all connections between the two clusters consisting of emerging connections. For each pair of clusters, we group the connections by topic, then examine how the connection between these two topics is being shaped by the emerging connections between low-level keywords.

\smallskip\noindent\textbf{RQ5 (Declining Keywords and Topics):} We address this question following a similar process to RQ4, based on \textit{oldest} average dates of publication. 66 keywords met the threshold in RQ4. Therefore, we also examine the 66 keywords with the oldest dates of publication. This corresponds to a period from November 2001--May 2011. 

These keywords represent concepts that are no longer receiving as much interest. This does not imply with certainty that such concepts are no longer relevant, or that they correspond to ``solved'' challenges. A keyword could represent (a) a topic or concept in decline (e.g., an older technology or approach that has been potentially superseded), (b) a well-established topic or concept with steady---but not growing---activity, or (c), a topic or concept that had a ``boom'' period in the past and a lower level of activity in recent years. Keywords may experience a resurgence. However, they have reduced relevancy to current development or testing trends, challenges, and research topics. 

Before stating that a particular topic is in decline, we compare the list of keywords and topics with those in RQ4. We say that a topic is declining in interest if both (a) it has several keywords with older average publication dates, \textbf{and} (b), lacks keywords with recent average dates. By examining both the oldest and newest keywords, we can more carefully discuss whether a topic is potentially in decline. 

\section{Results and Discussion}\label{sec:results}

Our analyses are based on 406 keywords, which are mapped into 11 clusters. We analyze this map by identifying the most popular keywords by occurrences and citations (Section~\ref{sec:topics_popular}) and the overarching research topics of each cluster (Section~\ref{sec:topics_clusters}), examining how keywords and topics are linked across clusters (Section~\ref{sec:across_clusters}), and exploring keywords, topics, and connections that are emerging or in potential decline (Sections~\ref{sec:topics_emerging}-\ref{sec:topics_declining}). We also offer advice and further exploratory analyses in Section~\ref{sec:advice}. 

\begin{figure}[!t]
    \centering
    \includegraphics[width=\textwidth]{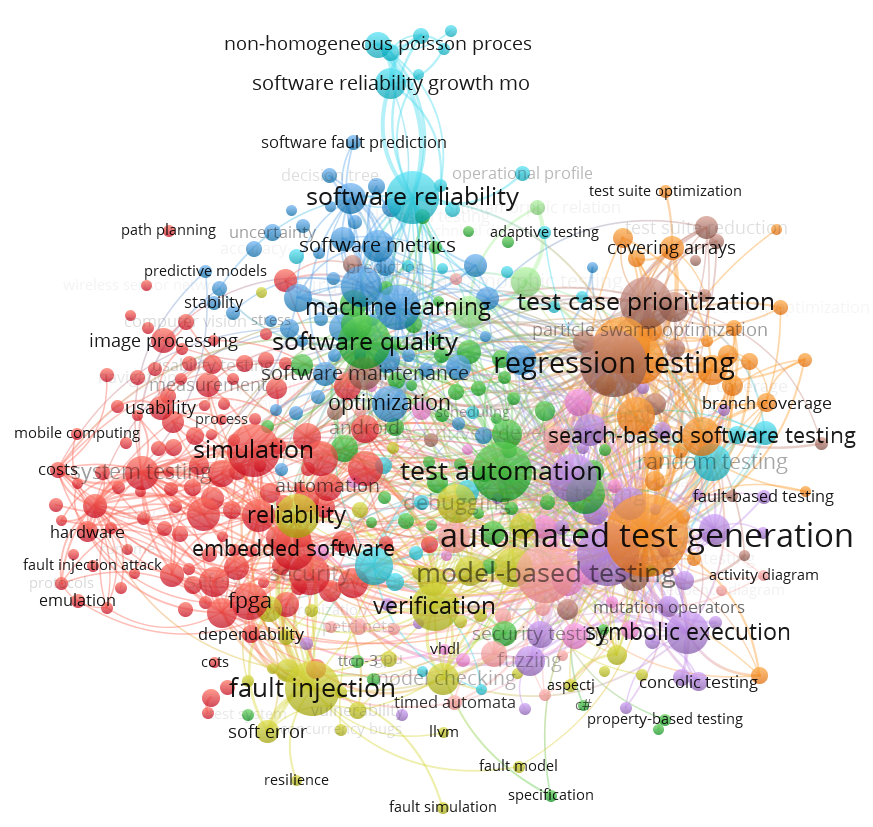}
    \caption{A visualization of the connections between publication keywords.}
    \label{fig:topics}
\end{figure}

\begin{center}
\begin{framed}
A visualization of the keyword map is shown in Figure~\ref{fig:topics}. An interactive version of this map can be accessed at \url{https://greg4cr.github.io/other/2021-TestingTrends/topics.html} or in the replication package.
\end{framed}
\end{center}

\subsection{RQ1: Popular Keywords}\label{sec:topics_popular}

\begin{table}[!t]
\resizebox{\columnwidth}{!}{
\begin{tabular}{llllll}
\hline
\textbf{Keyword} & \textbf{\# Pubs.} & \textbf{Percent} & \textbf{Citations} & \textbf{Date} & \textbf{Description} \\ \hline
Automated Test Generation & 1068 & 2.14\% & 16.36 & 2013 & The use of tools to generate full or partial test cases~\cite{Anand13:Orchestrated}. \\ \hline
Regression Testing & 701 & 1.41\% & 14.03 & 2014 &  \begin{tabular}[c]{@{}l@{}}A practice where tests are re-executed when code changes to \\ ensure that working code operates correctly~\cite{Korel98:ART}.\end{tabular} \\ \hline
Mutation Testing & 596 & 1.20\% & 14.83 & 2014 & \begin{tabular}[c]{@{}l@{}}A practice where synthetic faults 
are seeded into systems to \\ assess the sensitivity of tests~\cite{Just14:MMF}.\end{tabular} \\ \hline
Test Automation & 567 & 1.14\% & 9.71 & 2014 & Tools and practices that enable automation of test execution~\cite{Fewster99:Automation}.\\ \hline
Model-based Testing & 552 & 1.11\% & 8.38 & 2014 & \begin{tabular}[c]{@{}l@{}}Use of behavioral models to analyze the system, to design or \\ generate test cases, or to judge results of testing~\cite{Anand13:Orchestrated}.\end{tabular} \\ \hline
Genetic Algorithm & 519 & 1.10\% & 10.10 & 2014 & \begin{tabular}[c]{@{}l@{}}An optimization algorithm that models how populations evolve over \\ time~\cite{Mitchell98:IGA}. Often used to automate tasks.\end{tabular} \\ \hline
Fault Injection & 477 & 0.96\% & 6.12 & 2015 & Injection of faults into a system for analysis~\cite{Voas97:FaultInjection}. \\ \hline
Software Quality & 445 & 0.89\% & 8.14 & 2012 &  \begin{tabular}[c]{@{}l@{}}Means to define, measure, and assure the quality of software~\cite{Kitchenham96:Quality}. \\ Encompasses correctness and quality (e.g., performance or scalability). \end{tabular} \\ \hline
Simulation & 442 & 0.89\% & 4.53 & 2013 & \begin{tabular}[c]{@{}l@{}}Simulated execution of a system. May encompass how to simulate~\cite{Herzner05}, \\ testing in simulation~\cite{Mok96:SimulationVerification}, or obtaining realistic results~\cite{Gay16:SteeringTSE}. \end{tabular} \\ \hline
Software Reliability & 440 & 0.88\% & 12.71 & 2010 & \begin{tabular}[c]{@{}l@{}}Means to define, measure, and assess the how quality changes \\ over time~\cite{Crossley00:Quality}.\end{tabular} \\ \hline
Test Case Prioritization & 418 & 0.84\% & 17.42 & 2015 & \begin{tabular}[c]{@{}l@{}}Automated techniques that select a subset of tests for execution~\cite{Rothermel02:Reduction}.\end{tabular} \\ \hline
Verification & 366 & 0.73\% & 16.58 & 2012 & \begin{tabular}[c]{@{}l@{}}Techniques that assess whether software possesses a property of interest, \\ often using formal specifications~\cite{Pezze06:testing}. Testing is one verification technique.\end{tabular} \\ \hline
Coverage Criteria & 362 & 0.73\% & 13.21 & 2012 & \begin{tabular}[c]{@{}l@{}}Measurements used to assess the strength of a test suite based on how \\ tests exercise code elements~\cite{Gay15:risks}. \end{tabular} \\ \hline
Combinatorial Testing & 349 & 0.70\% & 14.35 & 2015 & \begin{tabular}[c]{@{}l@{}}A technique for generating or selecting test input, based on coverage  \\ of representative values~\cite{Nie11:Combinatorial}. \end{tabular} \\ \hline
Machine Learning & 326 & 0.65\% & 8.32 & 2017 & \begin{tabular}[c]{@{}l@{}}Algorithms that make inferences from patterns detected in data. Used in, \\ e.g., automation~\cite{Harman13:oraclesurvey}, predictive modeling~\cite{Turhan09:CCWC}, or evaluation~\cite{Gay19:fitness}. \end{tabular} \\ \hline
Reliability & 306 & 0.62\% & 13.95 & 2013 & \begin{tabular}[c]{@{}l@{}}Often a synonym for software quality, but can also refer to hardware \\ quality or a blend of hardware/software. \end{tabular}\\ \hline
Symbolic Execution & 295 & 0.59\% & 14.34 & 2014 &  \begin{tabular}[c]{@{}l@{}} Analyses where software is executed in an abstract form where \\ one symbolic input matches many real inputs~\cite{Cadar:2008:KLEE}.\end{tabular}\\ \hline
Embedded Software & 268 & 0.54\% & 4.86 & 2014 &  \begin{tabular}[c]{@{}l@{}}Complex self-contained hardware and software systems~\cite{Zander11:MBT}.\end{tabular}\\ \hline
Neural Networks & 266 & 0.53\% & 8.53 & 2015 & \begin{tabular}[c]{@{}l@{}}Network structures inspired by the human brain, used in \\ machine learning~\cite{Fukuda92}.\end{tabular}\\ \hline
Security & 265 & 0.52\% & 7.60 & 2015 &  \begin{tabular}[c]{@{}l@{}}Practices, tools, and techniques intended to prevent misuse of a system's \\ capabilities or data~\cite{vanLamsweerde07:kaos}.\end{tabular}\\ \hline
\end{tabular}}
\caption{
Keywords targeted in at least 0.50\% of publications ($\geq$ 250 publications). Each keyword is named and described, and the number of publications where the keyword is targeted, percentage of the sample, average date of publication, and average number of citations per study are included.
} 
\label{tab:popular_topics} \vspace{10pt}
\end{table}

We begin by identifying the most popular individual keywords, sorted by the number of publications (and percentage of the total sample). These keywords are listed in Table~\ref{tab:popular_topics}, along with a description, percentage of the sample, average age of publication (rounded to the year), and average number of citations per publication. 

These keywords are those that the authors considered important enough to note as one of the core focuses of their work. There are certainly more than 326 publications in this sample that \textit{use} machine learning, for example. However, the authors may not have listed machine learning as a keyword. Therefore, these keywords should be interpreted as the research concepts the authors felt were the most important and relevant.  

\begin{center}
\begin{framed}
\textbf{RQ1 (Popular Keywords):} The most common keywords tend to relate to automation, test creation and assessment guidance, assessment of system quality, and cyber-physical systems.
\end{framed}
\end{center}

Automation offers promise for increasing the quality and efficiency of testing, and many keywords (e.g., automated test generation, test automation) relate to automation. Additionally, genetic algorithms and symbolic execution often enable automation. Test case prioritization enables efficient test execution, and regression testing is a process performed as part of a test execution pipeline. Combinatorial testing suggests an important subset of test input to apply, often as part of automated generation. Models are often used to generate test input. Machine learning, including neural networks, supports prediction tasks related to automation.

\begin{table}[!t]
\resizebox{\columnwidth}{!}{
\begin{tabular}{llllll}
\hline
\textbf{Keyword} & \textbf{Citations} & \textbf{\# Pubs.} & \textbf{Date} & \textbf{Description} \\ \hline
Testing Strategies & 55.50 & 26 & 2010 &  \begin{tabular}[c]{@{}l@{}}Guiding principles for test design and the  testing process~\cite{jamil2016software}. \end{tabular}  \\ \hline
Testing and Debugging & 48.48 & 21 & 2013 &  \begin{tabular}[c]{@{}l@{}}
Debugging practices isolate and diagnose faults in the source code~\cite{Zeller02:Delta}.\\ This keyword relates to the combination of testing and debugging techniques.\end{tabular} \\ \hline
Partition Testing & 35.59 & 54 & 2005  &  \begin{tabular}[c]{@{}l@{}} 
Test input selection based on division of the system's input domain \\ into partitions, based on a set of rules~\cite{weyuker1991analyzing}. \end{tabular} \\ \hline
Fault-based Testing & 34.48 & 33 & 2005  &  \begin{tabular}[c]{@{}l@{}} 
Use of pre-specified faults in a program to create and evaluate test \\ suites~\cite{morell1990theory}. Mutation testing is an automated form of fault-based testing. \end{tabular}  \\ \hline
Constraints & 33.76 & 25 & 2011  &  \begin{tabular}[c]{@{}l@{}}Conditions that must be met to accomplish a goal, e.g., for input to take \\ a particular path in a program~\cite{Cadar:2008:KLEE}.\end{tabular} \\ \hline
Test Suite Minimization & 32.97 & 39 & 2014  & Process of reducing test suite size by eliminating redundant test cases~\cite{yoo2012regression}.  \\ \hline
Random Testing & 32.88 & 218 & 2011  & Testing software by generating random input~\cite{arcuri2011random}. \\ \hline
Software Fault Prediction & 31.82 & 38 & 2016  & Prediction of fault-prone code using  software metrics and fault metadata~\cite{catal2011software}. \\ \hline
Covering Arrays & 28.50 & 96 & 2013  & \begin{tabular}[c]{@{}l@{}}
The set of test specifications selected during combinatorial testing~\cite{Nie11:Combinatorial}. \end{tabular}  \\ \hline
Compiler Testing & 27.28 & 32 & 2014  & \begin{tabular}[c]{@{}l@{}}
Specialized testing practices for compilers, e.g., the selection of input\\  programs to ensure the compiler conforms to its target language's \\ semantics and syntax~\cite{chen2020survey}. \end{tabular} \\ \hline
Object Oriented Modeling & 25.75 & 20 & 2004  & \begin{tabular}[c]{@{}l@{}} Model formats based on object-oriented  design and object interaction~\cite{UMLReference}\end{tabular}  \\ \hline
Evolutionary Testing & 23.93 & 46 & 2011  & \begin{tabular}[c]{@{}l@{}}
The use of evolutionary algorithms (e.g., genetic algorithms) to generate test\\ input  or automate other tasks~\cite{Anand13:Orchestrated}. \end{tabular} \\ \hline
Test Design & 23.42 & 33 & 2013 & \begin{tabular}[c]{@{}l@{}} The process of defining test cases~\cite{copeland2004practitioner}. \end{tabular} \\ \hline
Regression Test Selection & 23.40 & 58 & 2014  & \begin{tabular}[c]{@{}l@{}}Practices to test cases for use during regression testing (e.g., only execute \\ tests for changed code)~\cite{rothermel1996analyzing}. \end{tabular} \\ \hline
Monte Carlo & 22.78 & 23 & 2015  & \begin{tabular}[c]{@{}l@{}}A family of algorithms used for optimization, numeric integration, and \\ probability assessment~\cite{ahmed2020evaluation}. \end{tabular} \\ \hline
Alloy & 22.18 & 22 & 2014  & Language for expressing complex behavior and constraints in software~\cite{jackson2019alloy}. \\ \hline
Automated Debugging & 21.76 & 38 & 2012  & Automated debugging techniques~\cite{zeller2001automated}. \\ \hline
Adaptive Random Testing & 21.51 & 95 & 2012 & \begin{tabular}[c]{@{}l@{}}Random testing techniques designed to ensure input is evenly spread over \\ the input domain\cite{chen2010adaptive}.
\end{tabular} \\ \hline
Data Flow Testing & 21.47 & 43 & 2011 & \begin{tabular}[c]{@{}l@{}}
Testing based on the flow of information between variable definitions\\ and usages~\cite{su2017survey}. \end{tabular} \\ \hline
Data Flow & 21.42 & 36 & 2008 & Metrics for tracking the flow of information~\cite{su2017survey} \\ \hline
Software Standards & 20.92 & 26 & 2009 & \begin{tabular}[c]{@{}l@{}}
Constraints, rules, and requirements that software or testing is  expected \\ to meet~\cite{DO178B}.\end{tabular}  \\ \hline
Synchronization & 20.07 & 29 & 2015 & \begin{tabular}[c]{@{}l@{}}Practices for ensuring components are able to coordinate when completing \\ tasks in parallel~\cite{hong2012testing}. \end{tabular} \\ \hline
Sensitivity Analysis & 20.00 & 36 & 2012 & \begin{tabular}[c]{@{}l@{}}Study of how uncertainty in system output can be traced to sources of \\ uncertainty in its inputs~\cite{douglas2020certain}. \end{tabular} \\ \hline
\end{tabular}}
\caption{
Keywords that received more than 20 citations on average per publication, with description, average number of citations, number of publications where the keyword is targeted, and average date of publication.
} 
\label{tab:popular_topics_citations}
\end{table}

Many of the remaining keywords relate to assessments of testing effectiveness or test creation guidance, including mutation testing, fault injection, and coverage criteria. Other keywords (software quality, reliability, verification, security) relate to the overall quality of the system, including its correctness, performance, and security. Finally, embedded software and simulation relate to systems combining software and hardware elements, which have high safety demands and unique testing activities~\cite{Gay16:SteeringTSE}. 

The average number of publications per keyword is 75 (0.15\%)---far below the number of publications targeting the top 20 keywords, indicating their importance. It is interesting that the most popular keyword is only a target of 2.14\% of the sample, and that only five keywords are targets of over 1\% of the sample. We believe this is an indication of the breadth of testing research. There are many challenges associated with testing, from test creation to automation, execution, assessment, and process. There are many ways to address each challenge, including algorithms and tools, human-driven activities, and studies of those working in the field. Even the median---40 publications---is a reasonable body of work on any single concept. 

We can compare the most-common keywords with the most-cited. In Table~\ref{tab:popular_topics_citations}, we identify keywords that receive, on average, the most citations per publication. 

\begin{center}
\begin{framed}
\textbf{RQ1 (Popular Keywords):} The most-cited keywords also relate to automation, test creation and assessment guidance, and assessment of system quality.
\end{framed}
\end{center}

Some of these keywords are linked to the most common keywords---fault-based testing, test suite minimization, covering arrays, partition testing, evolutionary testing, and regression test selection, in particular. However, no keywords appear in both lists. 

\begin{figure}[!t]
    \centering
    \includegraphics[width=0.8\textwidth]{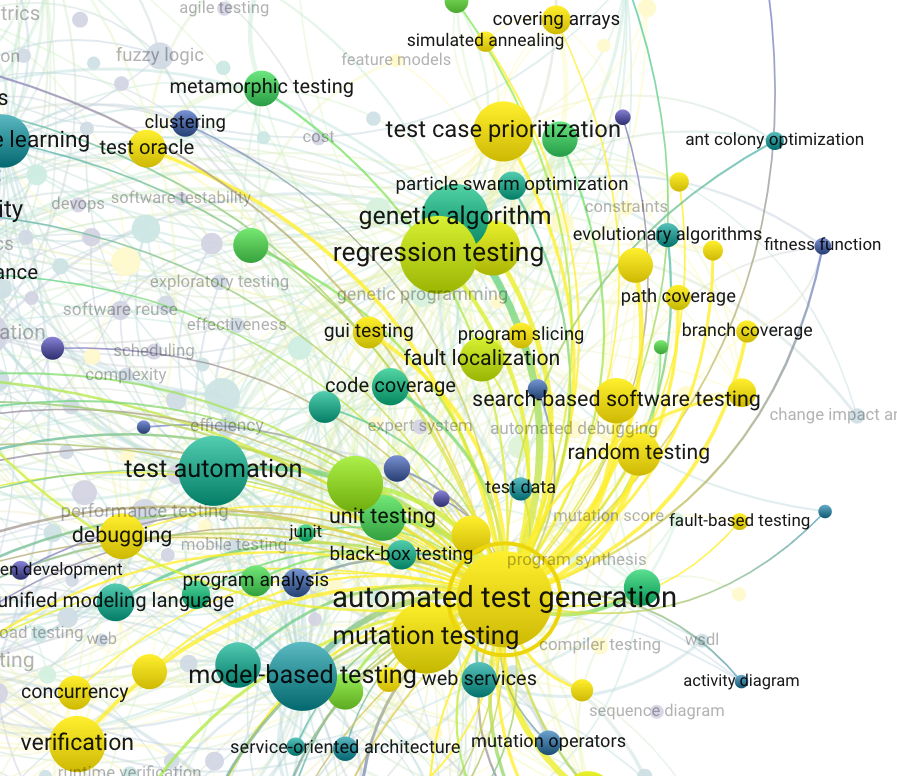}
    \caption{A subset of keywords connected to automated test generation, colored by the average number of citations. Nodes in yellow attract a high number of citations ($\geq 14$).}
    \label{fig:citations}
\end{figure}

However, both the most common and most-cited keywords share common themes. Many of the keywords relate to automation (e.g., test suite minimization, random testing), test creation and assessment (e.g., testing strategies, data flow testing), or quality assessment (e.g., software fault prediction, sensitivity analysis). For example, Figure~\ref{fig:citations} illustrates that many keywords associated with automated test generation receive a relatively high number of citations on average. 

In general, the keywords in Table~\ref{tab:popular_topics_citations} are associated with a relatively small number of publications. They also have an older average date of publication, approximately 2010 versus 2014, allowing more time to attract citations. We hypothesize that these particular keywords (a) are related to themes that attract attention, and (b) are attached to a small set of publications containing a subset of particularly influential publications.

\subsection{RQ2: Characterization of Clusters into Topics}\label{sec:topics_clusters}

\begin{figure}[!b]
        \centering
\includegraphics[width=\textwidth]{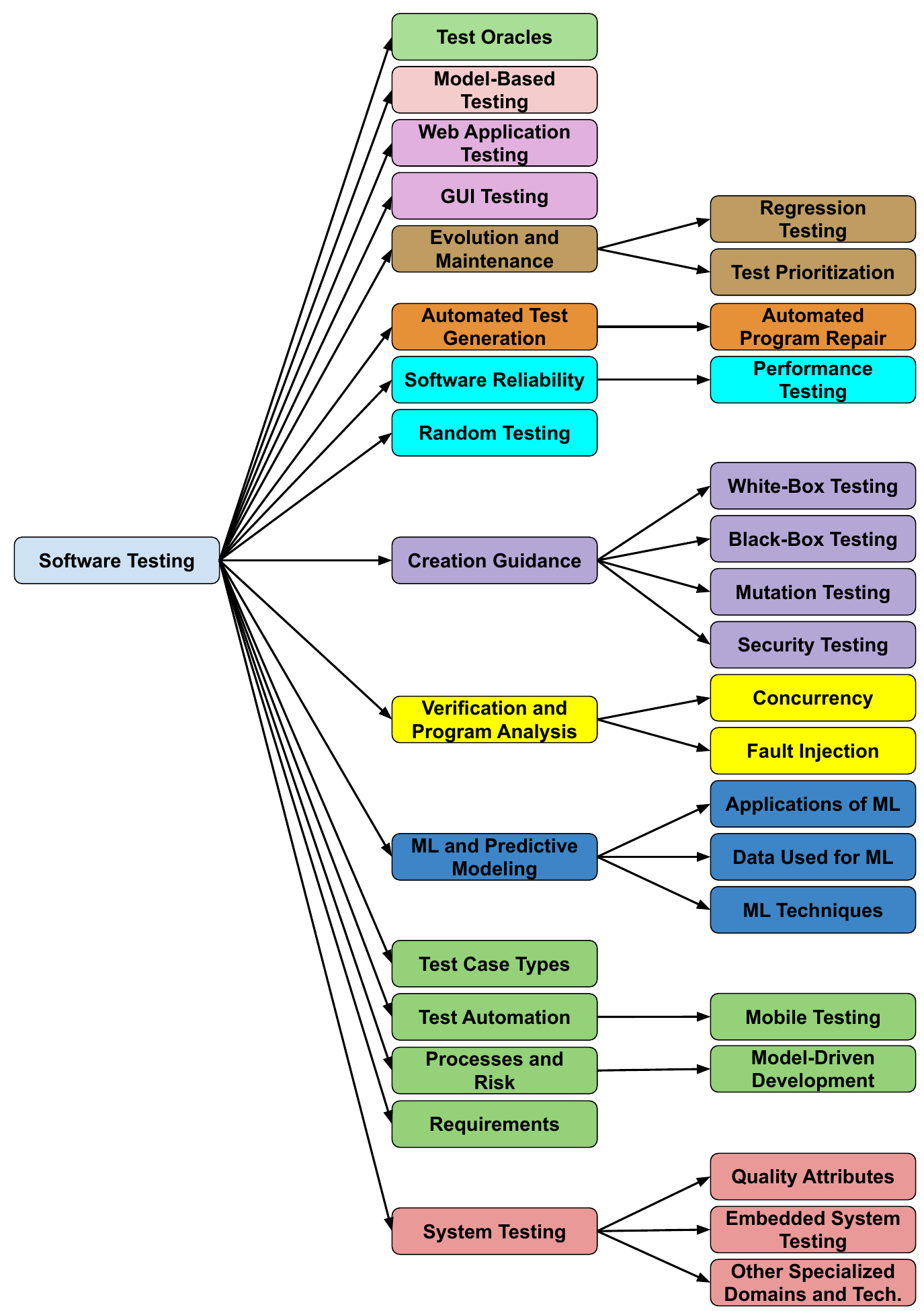}
  \caption{Identified research topics (middle layer) and subtopics (final layer), colored by cluster.}
    \label{fig:summary} 
\end{figure}

\begin{table}[!t]
\resizebox{\columnwidth}{!}{
\begin{tabular}{llllll}
\hline
\textbf{Cluster} & \textbf{Num} & \textbf{Density} & \textbf{Topics and} & \textbf{Example Keywords} & \textbf{Description} \\ 
& \textbf{Keywords} & & \textbf{\textit{(Subtopics)}} & & \\ \hline
11 & 4 & 100\% & \textbf{Test Oracles} & \begin{tabular}[c]{@{}l@{}}test oracle, \\ metamorphic relation\end{tabular} & \begin{tabular}[c]{@{}l@{}}Test component that issues a verdict of  correctness~\cite{Harman13:oraclesurvey}.\end{tabular} \\ \hline
10 & 16 & 39\% & \begin{tabular}[c]{@{}l@{}}\textbf{Model-Based}\\ \textbf{Testing}\end{tabular} & \begin{tabular}[c]{@{}l@{}}model transformation, \\ timed automata\end{tabular} & \begin{tabular}[c]{@{}l@{}}Use of behavioral models to analyze the system, \\ to design or generate test cases, or as oracles~\cite{Anand13:Orchestrated}.\end{tabular} \\ \hline
\multirow{3}{*}{9} & \multirow{3}{*}{18} & \multirow{3}{*}{31\%} &\textbf{Web Testing} &  \begin{tabular}[c]{@{}l@{}}web applications, \\ javascript\end{tabular} & \begin{tabular}[c]{@{}l@{}}Testing techniques, tools, and activities focused on \\ verification of web-based applications~\cite{Bozkurt10:Web}.\end{tabular} \\ \cline{4-6}
& & & \textbf{GUI Testing} &  \begin{tabular}[c]{@{}l@{}}graphical user interface, \\ finite state machine\end{tabular} & \begin{tabular}[c]{@{}l@{}}Test design or generation techniques focused on \\ exercising a system through its graphical interface~\cite{Qureshi13:gui}.\end{tabular} \\ \hline
\multirow{5}{*}{8} & \multirow{5}{*}{19} & \multirow{5}{*}{47\%} & \begin{tabular}[c]{@{}l@{}}\textbf{Evolution and}\\ \textbf{Maintenance}\end{tabular} & \begin{tabular}[c]{@{}l@{}}program comprehension,\\ change impact analysis\end{tabular} & \begin{tabular}[c]{@{}l@{}}Practices for controlling and maintaining quality \\ as the system changes over time~\cite{MurphyHill12:How}.\end{tabular} \\ \cline{5-6}
& & & \begin{tabular}[c]{@{}l@{}}(\textit{Regression Testing})\end{tabular} & \begin{tabular}[c]{@{}l@{}}regression testing, \\ regression test selection\end{tabular} & \begin{tabular}[c]{@{}l@{}}A practice where tests are re-executed when code changes \\ to ensure that working code operates correctly~\cite{Korel98:ART}.\end{tabular} \\ \cline{5-6}
& & & \begin{tabular}[c]{@{}l@{}}(\textit{Test Prioritization})\end{tabular} & \begin{tabular}[c]{@{}l@{}}test case prioritization, \\ test case selection\end{tabular} & \begin{tabular}[c]{@{}l@{}}Automated techniques that select a subset of tests \\ for execution~\cite{Rothermel02:Reduction}.\end{tabular} \\ \hline
\multirow{3}{*}{7} & \multirow{3}{*}{28} & \multirow{3}{*}{48\%} & \begin{tabular}[c]{@{}l@{}}\textbf{Automated Test} \\ \textbf{Generation}\end{tabular} & \begin{tabular}[c]{@{}l@{}}genetic algorithms,\\ branch coverage\end{tabular} & \begin{tabular}[c]{@{}l@{}}The use of tools to generate full or partial test cases~\cite{Anand13:Orchestrated}.\end{tabular} \\ \cline{5-6}
& & & \begin{tabular}[c]{@{}l@{}}(\textit{Automated Program}\\ \textit{Repair})\end{tabular} & \begin{tabular}[c]{@{}l@{}}fault localization, \\ genetic programming\end{tabular} & \begin{tabular}[c]{@{}l@{}}Automated generation of patches for faulty programs~\cite{Martinez17:APRD4J}.\end{tabular} \\ \hline
\multirow{5}{*}{6} & \multirow{5}{*}{30} & \multirow{5}{*}{24\%} &  \textbf{Reliability} & \begin{tabular}[c]{@{}l@{}}reliability growth, \\ quality control\end{tabular} & \begin{tabular}[c]{@{}l@{}}Means to define, measure, and assess the how quality \\ changes over time~\cite{Crossley00:Quality}.\end{tabular} \\ \cline{5-6}
& & & (\textit{Performance Testing}) & \begin{tabular}[c]{@{}l@{}}load testing, \\ cloud testing\end{tabular} & \begin{tabular}[c]{@{}l@{}}Testing to assess performance and scalability of a system \\ under different operating conditions~\cite{Moghadam19:RL}.\end{tabular} \\ \cline{4-6}
& & & \begin{tabular}[c]{@{}l@{}}\textbf{Random Testing}\end{tabular} & \begin{tabular}[c]{@{}l@{}}adaptive random testing, \\ statistical testing\end{tabular} & \begin{tabular}[c]{@{}l@{}}Generation of random input for various purposes \\ (e.g., assessing reliability or performance)~\cite{Anand13:Orchestrated}\end{tabular} \\ \hline
\multirow{9}{*}{5} & \multirow{9}{*}{32} & \multirow{9}{*}{30\%} & \begin{tabular}[c]{@{}l@{}}\textbf{Creation Guidance}\end{tabular} & \begin{tabular}[c]{@{}l@{}}certification, \\ test adequacy\end{tabular} & \begin{tabular}[c]{@{}l@{}}Guidance for how a tester might approach test design---e.g.,\\ goals, input selection, and assessing test strength.\end{tabular} \\ \cline{5-6}
& & & \begin{tabular}[c]{@{}l@{}}(\textit{White-Box Testing})\end{tabular} & \begin{tabular}[c]{@{}l@{}}coverage criteria, \\ data flow\end{tabular} & \begin{tabular}[c]{@{}l@{}}Test creation based on source code~\cite{Gay15:risks}.\end{tabular} \\ \cline{5-6}
& & & \begin{tabular}[c]{@{}l@{}}(\textit{Black-Box Testing})\end{tabular} & \begin{tabular}[c]{@{}l@{}}specification-based testing,\\ black-box testing\end{tabular} & \begin{tabular}[c]{@{}l@{}}Test creation based on requirements and \\ other documentation~\cite{Whalen06:CovMetrics-ReqBased}.\end{tabular} \\ \cline{5-6}
& & & \begin{tabular}[c]{@{}l@{}}(\textit{Mutation Testing})\end{tabular} & \begin{tabular}[c]{@{}l@{}}mutation score,\\ mutation operators \end{tabular} & \begin{tabular}[c]{@{}l@{}}Test creation based on synthetic faults \\ seeded into a system~\cite{Just14:MMF}\end{tabular} \\ \cline{5-6}
& & & \begin{tabular}[c]{@{}l@{}}(\textit{Security Testing})\end{tabular} & \begin{tabular}[c]{@{}l@{}}penetration testing, \\ software vulnerability \end{tabular} & \begin{tabular}[c]{@{}l@{}}Test creation to assess the ability of a system to \\ prevent exploitation of vulnerabilities~\cite{Takanen08:fuzzing}.\end{tabular} \\ \hline
\multirow{5}{*}{4} & \multirow{5}{*}{35} & \multirow{5}{*}{31\%} & \begin{tabular}[c]{@{}l@{}}\textbf{Verification} \\ \textbf{and Analysis}\end{tabular} & \begin{tabular}[c]{@{}l@{}}dynamic analysis, \\ static analysis \end{tabular} & \begin{tabular}[c]{@{}l@{}}Analyses performed to ensure that software possesses \\ properties of interest (e.g., correctness, resilience)~\cite{Pezze06:testing}.\end{tabular} \\ \cline{5-6}
& & & \begin{tabular}[c]{@{}l@{}}(\textit{Concurrency})\end{tabular} & \begin{tabular}[c]{@{}l@{}}parallelization, \\synchronization\end{tabular} & \begin{tabular}[c]{@{}l@{}}Analyses of programs that execute over parallel \\ threads or processes~\cite{Clarke86}.\end{tabular} \\ \cline{5-6}
& & & \begin{tabular}[c]{@{}l@{}}(\textit{Fault Injection})\end{tabular} & \begin{tabular}[c]{@{}l@{}}fault model, \\ fault tolerance\end{tabular} & \begin{tabular}[c]{@{}l@{}}Injection of faults into a system for  analysis~\cite{Voas97:FaultInjection}. \end{tabular} \\  \hline
\end{tabular}}
\caption{
An overview of clusters 4 - 11, including the cluster ID from VOSViewer, the number of keywords, inter-cluster connection density (percentage of possible connections between keywords), identified topics and subtopics, example keywords for each topic, and a brief description of each topic. Clusters are ordered from smallest to largest.
} 
\label{tab:clusters} \vspace{20pt}
\end{table}

\begin{table}[!t]
\resizebox{\columnwidth}{!}{
\begin{tabular}{llllll}
\hline
\textbf{Cluster} & \textbf{Num} & \textbf{Density} & \textbf{Topics and} & \textbf{Example Keywords} & \textbf{Description} \\ 
& \textbf{Keywords} & & \textbf{\textit{(Subtopics)}} & & \\ \hline
\multirow{7}{*}{3} & \multirow{7}{*}{48} & \multirow{7}{*}{26\%} & \begin{tabular}[c]{@{}l@{}}\textbf{Machine Learning}\end{tabular} & \begin{tabular}[c]{@{}l@{}}machine learning\end{tabular} & \begin{tabular}[c]{@{}l@{}}Algorithms that make inferences from patterns \\  detected in data~\cite{Turhan09:CCWC}.\end{tabular} \\ \cline{5-6}
& & & \begin{tabular}[c]{@{}l@{}}(\textit{Applications})\end{tabular} & \begin{tabular}[c]{@{}l@{}}defect prediction, \\ estimation\end{tabular} & \begin{tabular}[c]{@{}l@{}}Applications of ML in software testing.\end{tabular} \\ \cline{5-6}
& & & \begin{tabular}[c]{@{}l@{}}(\textit{Data Used})\end{tabular} & \begin{tabular}[c]{@{}l@{}}metrics, \\ complexity\end{tabular} & \begin{tabular}[c]{@{}l@{}}Sources of data used to draw conclusions with ML. \end{tabular} \\  \cline{5-6}
& & & \begin{tabular}[c]{@{}l@{}}(\textit{ML Techniques})\end{tabular} & \begin{tabular}[c]{@{}l@{}}neural networks, \\ deep learning\end{tabular} & \begin{tabular}[c]{@{}l@{}}ML techniques used in testing research.\end{tabular} \\  \hline
\multirow{11}{*}{2} & \multirow{11}{*}{58} & \multirow{11}{*}{21\%} & \begin{tabular}[c]{@{}l@{}}\textbf{Test Case Types}\end{tabular} & \begin{tabular}[c]{@{}l@{}} unit testing,\\ exploratory testing\end{tabular} & \begin{tabular}[c]{@{}l@{}}Practices and levels of granularity for test design.\end{tabular} \\ \cline{4-6}
& & & \begin{tabular}[c]{@{}l@{}}\textbf{Test Automation}\end{tabular} & \begin{tabular}[c]{@{}l@{}} test execution,\\ testing tools \end{tabular} & \begin{tabular}[c]{@{}l@{}}Tools and practices that enable automation of \\ test execution~\cite{Fewster99:Automation}. \end{tabular} \\ \cline{5-6}
& & & \begin{tabular}[c]{@{}l@{}}(\textit{Mobile Testing})\end{tabular} & \begin{tabular}[c]{@{}l@{}}mobile testing, \\ android testing\\ \end{tabular} & \begin{tabular}[c]{@{}l@{}}Testing techniques, tools, and activities focused on \\ verification of mobile applications~\cite{Alshahwan18:Sap}.\end{tabular} \\ \cline{4-6}
& & & \begin{tabular}[c]{@{}l@{}}\textbf{Processes and Risk}\end{tabular} & \begin{tabular}[c]{@{}l@{}} software quality,\\ test-driven development\end{tabular} & \begin{tabular}[c]{@{}l@{}} The organization, management, and testing process \\ of a development team~\cite{Pezze06:testing}. \end{tabular} \\ \cline{5-6}
& & & \begin{tabular}[c]{@{}l@{}}(\textit{Model-Driven} \\ \textit{Development})\end{tabular} & \begin{tabular}[c]{@{}l@{}}model-driven development, \\ model-driven testing\end{tabular} & \begin{tabular}[c]{@{}l@{}}Development process based on use of models for \\  analysis, code generation, and testing~\cite{FoSE07:Model-Based}\end{tabular} \\ \cline{4-6}
& & & \begin{tabular}[c]{@{}l@{}}\textbf{Requirements} \\ \textbf{Engineering}\end{tabular} & \begin{tabular}[c]{@{}l@{}} requirements engineering, \\ traceability\end{tabular} & \begin{tabular}[c]{@{}l@{}}Requirements indicate correct behavior. Verification \\ often assesses conformance of code to requirements~\cite{Pezze06:testing}.\end{tabular} \\ \hline
\multirow{8}{*}{1} & \multirow{8}{*}{118} & \multirow{8}{*}{20\%} & \begin{tabular}[c]{@{}l@{}}\textbf{System Testing}\end{tabular} & \begin{tabular}[c]{@{}l@{}}system testing, \\ user interfaces\end{tabular} & \begin{tabular}[c]{@{}l@{}}Test cases that interact with an external system interface~\cite{Pezze06:testing}.\end{tabular} \\ \cline{5-6}
& & & \begin{tabular}[c]{@{}l@{}}(\textit{Quality Attributes})\end{tabular} & \begin{tabular}[c]{@{}l@{}}usability, \\ software performance\end{tabular} & \begin{tabular}[c]{@{}l@{}}Non-functional properties of a system assessed as \\ part of quality assurance~\cite{Clements98:SoftArch} \end{tabular} \\ \cline{5-6}
& & & \begin{tabular}[c]{@{}l@{}}(\textit{Embedded Systems})\end{tabular} & \begin{tabular}[c]{@{}l@{}}real-time system, \\ simulation\end{tabular} & \begin{tabular}[c]{@{}l@{}}Complex self-contained hardware and software systems~\cite{Zander11:MBT}.\end{tabular} \\  \cline{5-6}
& & & \begin{tabular}[c]{@{}l@{}}(\textit{Other Specialized} \\ \textit{Domains})\end{tabular} & \begin{tabular}[c]{@{}l@{}}open source software, \\ image processing, \\ autonomous vehicles\end{tabular} & \begin{tabular}[c]{@{}l@{}}System types (e.g., databases, virtual reality, operating \\ systems) or technologies (e.g., XML, Java) with \\ dedicated testing approaches. \end{tabular} \\  \hline
\end{tabular}}
\caption{
An overview of clusters 1 - 3, including the cluster ID from VOSViewer, the number of keywords, inter-cluster connection density (percentage of possible connections between keywords), identified topics and subtopics, example keywords for each topic, and a brief description of each topic. Clusters are ordered from smallest to largest.
} 
\label{tab:clusters2} 
\end{table}

By examining the connections between keywords, we can understand the context in which keywords form, grow, and thrive. Therefore, we have identified \textit{research topics} characterizing the keyword clusters. These topics are detailed in Tables~\ref{tab:clusters}-\ref{tab:clusters2}. We note a cluster ID assigned by VOSviewer, the number of keywords in that cluster, the density of connections between keywords within each cluster (the ratio of the number of existing within-cluster connections to the total possible within-cluster connections, $\text{Keywords} \choose 2$), and the topics and subtopics assigned to that cluster. For each topic or subtopic, we list two example keywords that fall within that topic, and we briefly describe the meaning of the topic. For additional clarity, Figure~\ref{fig:summary} outlines these topics, colored by the cluster they emerged from. 

\begin{center}
\begin{framed}
\textbf{RQ2 (Characterization of Clusters into Topics):} Based on keyword clustering, testing research can be divided into 16 topics, with a further 18 subtopics.
\end{framed}
\end{center}



While some of the topics within a cluster may seem independent, they are linked by connections between the underlying keywords. It is important, therefore, to examine both topics and keywords to come to a full understanding of a particular cluster. For example, random testing is a topic with widespread applicability. However, it is linked to Cluster 6 because random testing is often used to assess reliability or performance.

Within Cluster 2, the test automation topic encapsulates the emerging subtopic of mobile testing. Mobile testing is not as well-established as web application testing, but is clearly growing as a distinct research area. In the future, it may emerge as an independent research topic---perhaps even as its own cluster. Additionally, the model-driven development subtopic in Cluster 2 is related to---but also separate from---the model-based testing topic in Cluster 10. The latter focuses on technical aspects of modelling, while the former focuses on process and practices that may use these technologies. There are connections between the two, but they contain different keywords.

Cluster 1 is the least cohesive cluster. However, we can categorize many keywords under a core topic of system testing. In Cluster 2, there is a topic centered around test case types (e.g., unit testing). System testing is often grouped with these test case types. However, it is also a broader concept encompassing many different types of systems and system interfaces (e.g., embedded systems, operating systems, or databases). Several topics in our characterization also relate to system-level practices or domains, e.g., web, GUI, and performance testing. Those topics are established enough to stand independently, while the system testing topic in Cluster 1 acts as a broad umbrella. 



\subsection{RQ3: Connections Across Clusters}\label{sec:across_clusters}

\begin{table}[!t]
\centering
\scriptsize 
\begin{tabular}{l|ccccccccccc}
\hline
\textbf{Cluster}   & \textbf{1}             & \textbf{2}             & \textbf{3}             & \textbf{4}             & \textbf{5}             & \textbf{6}             & \textbf{7}             & \textbf{8}             & \textbf{9}             & \textbf{10}            & \textbf{11}            \\ \hline
1  & \textit{0.20} & 0.08          & 0.09          & 0.11 & 0.06          & 0.06          & 0.06          & 0.07          & 0.07          & 0.09          & 0.07          \\
2  &               & \textit{0.21} & 0.08          & 0.09          & 0.10          & 0.08          & 0.09          & \textbf{0.13} & 0.10          & \textbf{0.14} & 0.11 \\
3  &               &               & \textit{0.26} & 0.10          & 0.08          & 0.08          & \textbf{0.12} & \textbf{0.13} & 0.05          & 0.07          & \textbf{0.15} \\
4  &               &               &               & \textit{0.31} & \textbf{0.15} & 0.08          & \textbf{0.12} & \textbf{0.12} & 0.07          & 0.10 & \textbf{0.12} \\
5  &               &               &               &               & \textit{0.30} & 0.08          & \textbf{0.20} & \textbf{0.15} & \textbf{0.13} & \textbf{0.14} & \textbf{0.21} \\
6  &               &               &               &               &               & \textit{0.24} & 0.10          & 0.11 & 0.06          & 0.06          & \textbf{0.12} \\
7  &               &               &               &               &               &               & \textit{0.48} & \textbf{0.23} & \textbf{0.14} & \textbf{0.14} & \textbf{0.15} \\
8  &               &               &               &               &               &               &               & \textit{0.47} & \textbf{0.13} & \textbf{0.15} & \textbf{0.20} \\
9  &               &               &               &               &               &               &               &               & \textit{0.31} & 0.10          & \textbf{0.17} \\
10 &               &               &               &               &               &               &               &               &               & \textit{0.39} & 0.09          \\
11 &               &               &               &               &               &               &               &               &               &               & \textit{1.00} \\ \hline
\end{tabular}
\caption{
Connection density between pairs of clusters. Cross-cluster densities $\geq 0.12$ are highlighted. Densities in italics represent within-cluster densities for each cluster.
} 
\label{tab:density} 
\end{table}

We analyze connections \textit{across} clusters by measuring the connection density between pairs of clusters, and by identifying keywords that bridge clusters. 

\smallskip\noindent\textbf{Connection Density:} Table~\ref{tab:density} shows all cross-cluster densities, with those $\geq 12\%$ are highlighted. 23 of the 45 pairs of clusters meet this threshold, indicating that many research topics are densely connected. 

\begin{center}
\begin{framed}
\textbf{RQ3 (Connections Across Clusters):} Clusters 5 (creation guidance), 7 (automated test generation), 8 (evolution and maintenance), and 11 (test oracles) are densely connected to several clusters. These clusters represent particularly multidisciplinary topics. 
\end{framed}
\end{center}

Cluster 8 appears in eight pairs, Clusters 7 and 11 appear in seven, and Cluster 5 appears in six. In particular, the pairings between Clusters 7 and 8 and between Clusters 5 and 11 have a higher connection density than the within-cluster densities of Clusters 1 and 2, indicating the dense interconnection between these topics.

As the most common keyword identified in our analysis, automated test generation has connections to keywords and topics in every other cluster. If a testing method exists, there will be an interest in generating tests for it (e.g., Clusters 5, 8, 9, and 10). Oracle creation often requires manual effort, leading to an interest in automated generation or reuse of oracles (Cluster 11). Further, machine learning offers the means to assist or enable automated test generation (Cluster 3). 

Test oracles are a necessary component of almost all test cases, leading to dense connections with Clusters 5 (creation guidance), 6 (reliability, performance, random testing), 8 (regression testing), and 9 (web and GUI testing). In addition to the above-mentioned connection to automated generation, machine learning offers a means to automate the creation of oracles (Cluster 3). Oracles are also a natural part of verification and different program analyses (Cluster 4). 

Maintenance has implications on multiple aspects of testing, such as costs and quality. Maintenance needs affect the tasks performed during test automation (Cluster 2). Test prioritization also uses the same information that guides test creation to select tests (Cluster 5), and can be assisted using machine learning (Cluster 3). Both regression testing and test prioritization are performed for GUIs and web applications (Cluster 9), and can make use of models (Cluster 10). Further, analyses related to program and test evolution are often connected to other analyses in Cluster 4. 

Test creation practices (Cluster 5) also connect broadly. Beyond automated test generation, test oracles, and test prioritization, several creation practices either have adaptations for model-based testing (Cluster 10) or for web and GUI testing (Cluster 9). In addition, there are connections between verification and test creation practices (Cluster 4)---e.g., black box testing and verification are connected through specifications, and security testing and analysis are related. 

Clusters 1, 2, 3, and 6 have the least-dense connections to other clusters. Clusters 1 and 2 are both large clusters with multiple topics and subtopics that are distinct, but closely-related. Connections exist to other clusters, but may be less common, as these two clusters already represent a broad set of keywords. Reliability and performance testing (Cluster 6) and various forms of predictive modelling in Cluster 3 are also often pursued as standalone topics, but can be connected to other topics. Out of all density measurements, the lowest was between Cluster 3 (machine learning) and Cluster 9 (web and GUI testing), with 5\% of possible connections existing in publications. 

\begin{table}[!t]
\resizebox{\columnwidth}{!}{
\begin{tabular}{lllll}
\hline
\textbf{Keyword} & \textbf{Connected} & \textbf{Connected} & \textbf{Position in} & \textbf{Description} \\ 
& \textbf{(External)} & \textbf{(All)} & \textbf{Table 1} & \\ \hline
Automated Test Generation & 217 & 243 & 1 & See Table~\ref{tab:popular_topics}. \\ \hline
Software Quality & 164 & 202 & 8 & See Table~\ref{tab:popular_topics}. \\ \hline
Mutation Testing & 164 & 184 & 3 & See Table~\ref{tab:popular_topics}. \\ \hline
Regression Testing & 157 & 174 & 2 & See Table~\ref{tab:popular_topics}. \\ \hline
Test Automation & 152 & 200 & 4 & See Table~\ref{tab:popular_topics}. \\ \hline
Model-based Testing & 150 & 163 & 5 & See Table~\ref{tab:popular_topics}. \\ \hline
Coverage Criteria & 147 & 168 & 13 & See Table~\ref{tab:popular_topics}. \\ \hline
Verification & 143 & 164 & 12 & See Table~\ref{tab:popular_topics}. \\ \hline
Genetic Algorithm & 139 & 161 & 6 & See Table~\ref{tab:popular_topics}. \\ \hline
Machine Learning & 132 & 161 & 15 & See Table~\ref{tab:popular_topics}. \\ \hline
Test Case Prioritization & 119 & 134 & 11 & See Table~\ref{tab:popular_topics}. \\ \hline
Software Maintenance & 119 & 130 & - & \begin{tabular}[c]{@{}l@{}}Practices for controlling and maintaining quality \\ as the system changes over time~\cite{MurphyHill12:How}.\end{tabular}\\ \hline 
Debugging & 117 & 135 & - & See Table~\ref{tab:popular_topics_citations}.\\ \hline 
Unit Testing & 114 & 136 & - & \begin{tabular}[c]{@{}l@{}}A practice where tests are created for a small,\\ isolated unit of code (typically a class)~\cite{Pezze06:testing}.\end{tabular} \\ \hline
Software Reliability & 114 & 132 & 10 & See Table~\ref{tab:popular_topics}. \\ \hline
Reliability & 108 & 125 & 16 & See Table~\ref{tab:popular_topics}. \\ \hline
Fault Injection & 106 & 128 & 7 & See Table~\ref{tab:popular_topics}. \\ \hline
Static Analysis & 101 & 120 & - & \begin{tabular}[c]{@{}l@{}}Analyses performed without executing the code \\ (e.g., inspection or symbolic execution)~\cite{Pezze06:testing}.\end{tabular}\\ \hline
Mutation Analysis & 101 & 116 & - & \begin{tabular}[c]{@{}l@{}}Analyses of programs or tests performed\\ using injected mutations~\cite{Just11:Major}.\end{tabular}\\ \hline
Unified Modeling Language & 101 & 111 & - & \begin{tabular}[c]{@{}l@{}}A family of techniques for modelling and \\analyzing program behavior~\cite{UMLReference}.\end{tabular}\\ \hline
\end{tabular}}
\caption{
Keywords that are connected to at $\geq 100$ keywords in clusters other than the one where the keyword is assigned (both keywords are targeted in at least one study). Each keyword is named and described, and the number of connected keywords (in external clusters, and in total) are listed. 
} 
\label{tab:connecting_keywords} 
\end{table}

\smallskip\noindent\textbf{Connecting Keywords:} In Table~\ref{tab:connecting_keywords}, we list all keywords that are connected to at least 100 keywords in external clusters. 

\begin{center}
\begin{framed}
\textbf{RQ3 (Connections Across Clusters):} Twenty keywords serve as ``connectors'' between clusters, reflecting multidisciplinary concepts (e.g., software quality), common test activities (e.g., unit testing), and common sources of information for test creation (e.g., coverage criteria). 
\end{framed}
\end{center}

For comparison, we also list the total number of connected keywords, and the position that the keyword had in Table~\ref{tab:popular_topics} (if it appeared in the most commonly-targeted keywords). Many of the connecting keywords are also among the most common occurring keywords, with automated test generation on top of both lists. The exact positions of keywords shift in the ordering, but 14 of the 20 most common keywords are also connecting keywords. The most common keywords tended to relate to automation, test creation and assessment guidance, assessment of system quality, and cyber-physical systems. These concepts---especially the first three---are broad, with wide-ranging applicability. That suggests that popularity of a keyword is not only a reflection of a particular concept, but on its multidisciplinary applicability.

In contrast to Table 1, we see a notable rise in the position of software quality, coverage criteria, and machine learning. Software quality and machine learning are both very broad concepts, while coverage criteria are a common source of information and a target for testing, with applications in test creation guidance, automated test generation, quality assessment, prediction, and other areas. 

We also see several keywords emerge: software maintenance, debugging, unit testing, static analysis, mutation analysis, and unified modeling language. These include broadly applicable concepts (maintenance, debugging, static analysis, mutation analysis), a common source of information (unified modeling language), and a common testing activity (unit testing). 

Six of the most common keywords do not meet the threshold for connecting keywords---simulation (93 external connections), combinatorial testing (96), symbolic execution (83), embedded software (85), neural networks (83), and security (82). All six are multidisciplinary concepts, but are more specific---rather than broad---concepts (combinatorial testing, symbolic execution, embedded software, neural networks).  

\begin{figure}[!t]
    \centering
    \includegraphics[width=\textwidth]{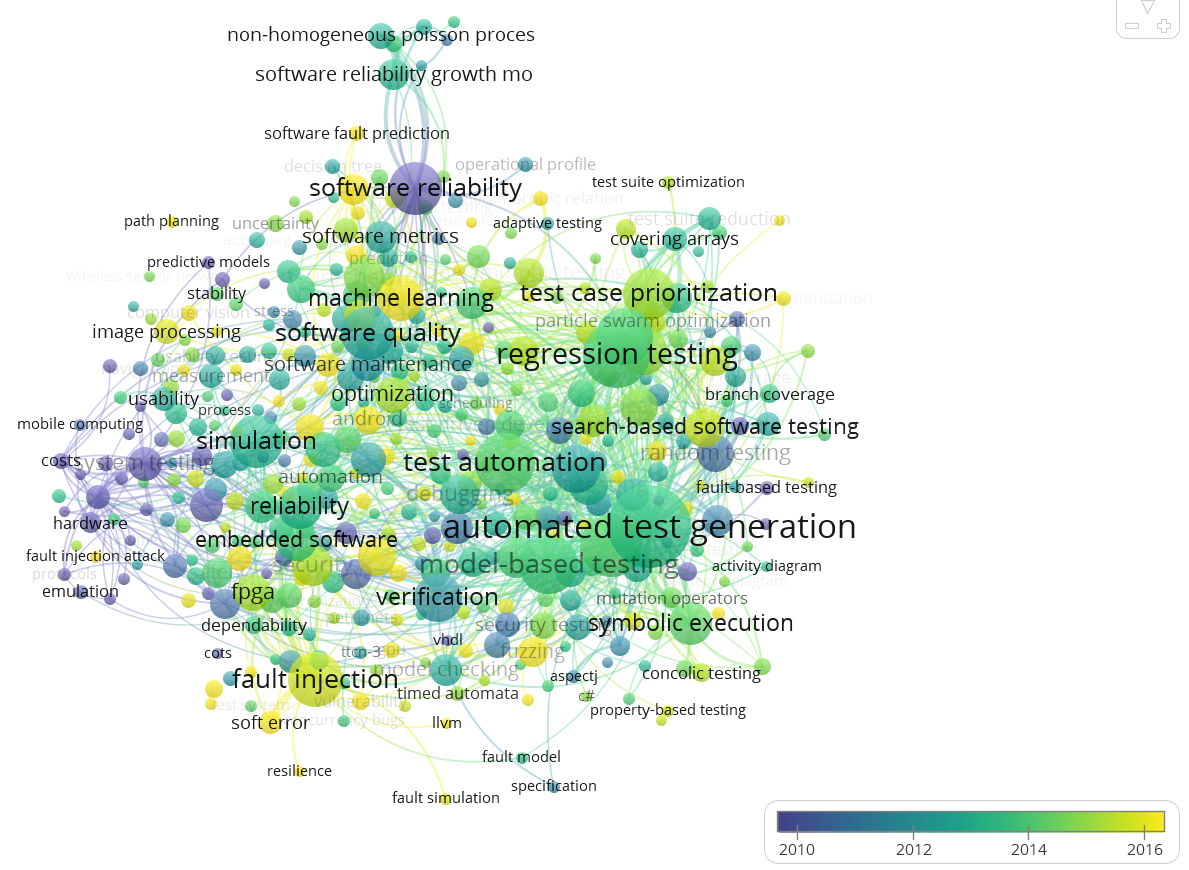}
    \caption{The map of keywords, colored by the average year of publication. Note that ``2010'' should be read as $\leq 2010$ and ``2016'' should be read as $\geq 2016$.}
    \label{fig:topics_historical}
\end{figure}

\subsection{RQ4: Emerging Keywords, Topics, and Connections}\label{sec:topics_emerging}

A visualization of the map of keywords, colored by average year of publication, is shown in Figure~\ref{fig:topics_historical}. Yellow nodes have an average date of 2016 or newer. Blue nodes have an average date of 2010 or earlier. A gradient between blue and yellow represents 2010--2016. We examine keywords, topics, and connections that have emerged or grown in interest since June 2015.

\begin{center}
\begin{framed}
An interactive version of this map can be accessed as an overlay at \url{https://greg4cr.github.io/other/2021-TestingTrends/topics.html} by selecting ``Avg. Pub. Year'' under the ``Color'' option.
\end{framed}
\end{center}

\begin{figure}[!t]
        \centering
\includegraphics[width=0.85\textwidth]{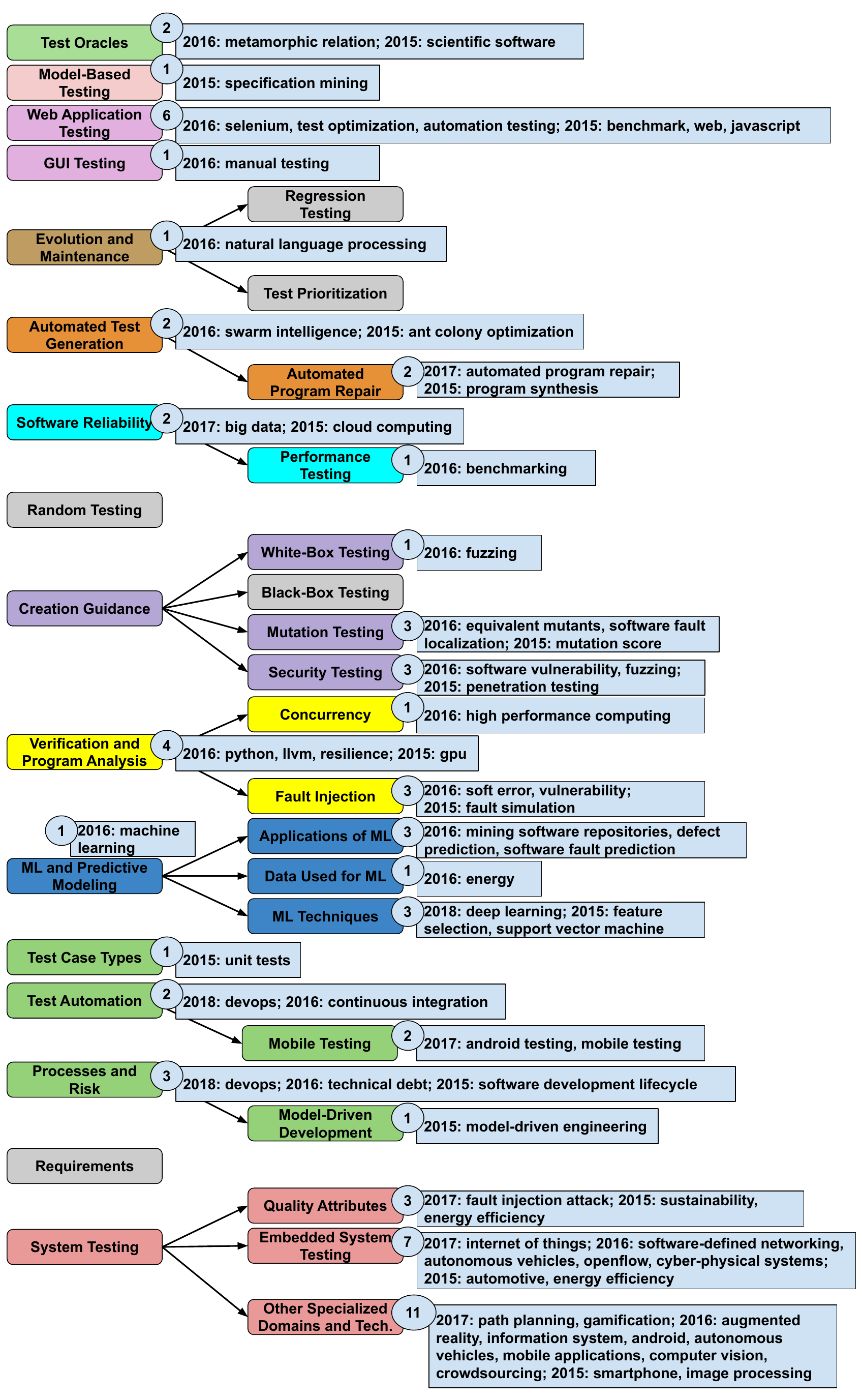}
  \caption{Keywords with an average publication date \textbf{newer than June 2015}, along with their associated research topic. The number next to the list of keywords indicates the number of emerging keywords. Topics colored in gray are those without emerging keywords.}
    \label{fig:growth_areas} \vspace{-25pt}
\end{figure}

\smallskip\noindent\textbf{Keywords and topics:} Sixty-six keywords (16.26\%) represent new emerging concepts or have received significant recent attention. Figure~\ref{fig:growth_areas} links these keywords to their respective research topic.
From these results, we can make several observations:
\begin{itemize}
\setlength{\itemsep}{1pt}
  \setlength{\parskip}{0pt}
  \setlength{\parsep}{0pt} 
    \item Many of the growth areas map to shifts in technology. There is growing interest in web applications, relating to technologies (JavaScript), testing tools (Selenium), and testing techniques. There is a similar emergence of mobile applications, in both the subtopic of mobile testing in Cluster 2 (android testing, mobile testing) and technologies in Cluster 1 (mobile applications, smartphone). 
    \item Machine learning has advanced many fields.  Unsurprisingly, it is also one of the largest growth areas in testing. The keyword ``machine learning'' has an average publication date of October 2016, and keywords have emerged related to applications, data, and specific techniques for ML. ``Deep learning'' is one of the newest keywords (average date of September 2018). 
    \item Keywords have also emerged targeting ML and AI-based systems. From the embedded systems and ``other domains'' topics, we see keywords related to autonomous vehicles, computer vision, image processing, and augmented reality. All of these areas require specialized testing approaches. 
    Autonomous vehicles, in particular, may grow into its own independent subtopic in the future.
    \item There is growing interest in energy consumption. This is connected to mobile applications, and a shift to portable devices that rely on batteries. This also reflects growing interest in sustainability and environmental impact of software.
    \item Automated program repair has emerged as a subtopic. The core keyword has one of the newest average publication dates (March 2017), and its connected keywords (e.g., program synthesis) also have recent dates. 
    \item Fuzzing and search-based approaches (swarm intelligence, ant colony optimization) have emerged as test generation techniques. Fuzzing, notably, has seen application in general and security-focused testing topics. Security-related keywords are also active and growing. 
\end{itemize}

\begin{center}
\begin{framed}
\textbf{RQ4 (Emerging Keywords, Topics, and Connections):} Emerging keywords and topics relate to, or incorporate, web and mobile applications, machine learning and AI---including autonomous vehicles---energy consumption, automated program repair, or fuzzing and search-based test generation.
\end{framed}
\end{center}

\begin{figure}[!t]
        \centering
\includegraphics[width=0.95\textwidth]{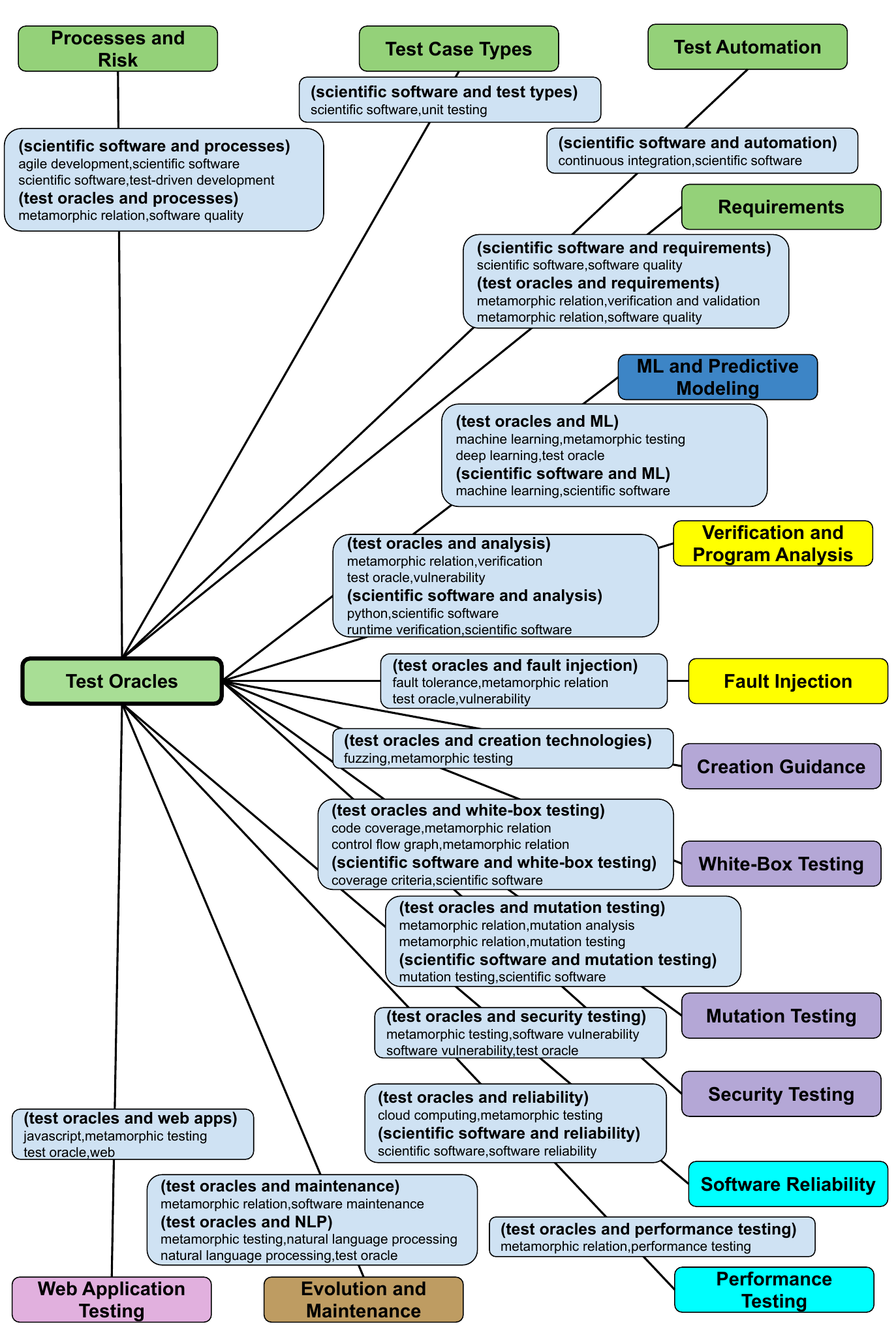} 
  \caption{Emerging connections, connected by research topic with test oracles, for the cluster pairings with highest ratio of emerging to total connections.} \vspace{-10pt}
    \label{fig:emerging_connections} 
\end{figure}

\begin{figure}[!t]
        \centering
\includegraphics[width=0.95\textwidth]{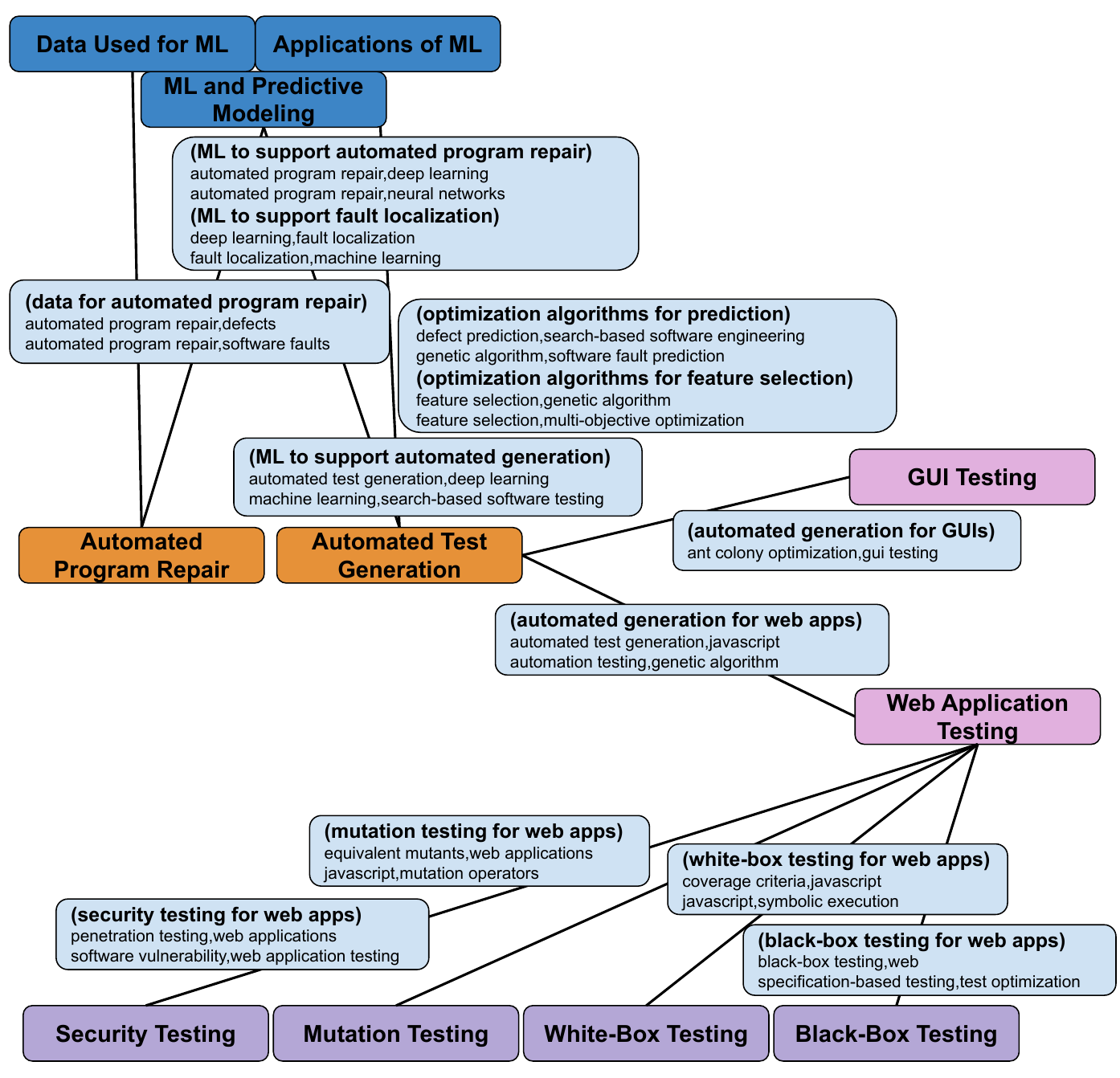} 
  \caption{Emerging connections, connected by research topic (excluding test oracles), for the cluster pairings with highest ratio of emerging to total connections.} \vspace{-10pt}
    \label{fig:emerging_connections_2} 
\end{figure}

\smallskip\noindent\textbf{Connections:} We focus our examination on ten pairs of clusters with the highest proportion of emerging connections to the number of possible connections ($\geq 3.5\%$). The connected clusters, and their associated topics, have a rapidly evolving relationship. 
\begin{itemize}
    \item Cluster 11 (test oracles) with Clusters 5 (creation guidance; 8.59\% of connections are emerging), 3 (machine learning; 6.77\%), 8 (evolution and maintenance; 5.26\%), 6 (reliability; 4.17\%), 9 (web application and GUI testing; 4.17\%), 2 (test case types, test automation, processes and risk, and model-driven development; 3.88\%), and 4 (verification and program analysis; 3.57\%).
    \item Cluster 7 (automated test generation) with Clusters 9 (4.36\%) and 3 (3.57\%). 
    \item Cluster 5 with Cluster 9 (4.69\%). 
\end{itemize}
The highlighted connections between topics are shown in Figure~\ref{fig:emerging_connections} for topics connected with test oracles, and in Figure~\ref{fig:emerging_connections_2} for other topics. For each connection between topics, a small number of example connections between keywords are shown. 

\begin{center}
\begin{framed}
\textbf{RQ4 (Emerging Keywords, Topics, and Connections):} Web applications and scientific computing require targeted testing approaches and practices, leading to emerging connections to many topics. Test oracles are also a rapidly-evolving topic with many emerging connections. Machine learning has emerging potential to support automation.
\end{framed}
\end{center}

\noindent We make several observations about these emerging connections:
\begin{itemize}
    \item Test oracles appear often because (a) Cluster 11 is a small cluster, (b) this topic has the largest percentage of emerging keywords, and (c), this topic is naturally connected to all other topics. Research interest in test oracles is growing~\cite{Harman13:oraclesurvey,Fontes21:oracle}, and effective oracles are needed for emerging domains such as web applications. The relationship between oracles and different testing practices is not well understood yet, leading to many emerging connections. Further, interest is growing in the use of machine learning to generate test oracles~\cite{Fontes21:oracle}. 
    \item The keyword ``scientific computing'' is part of Cluster 11, due to its frequent connection with metamorphic testing. Inspection of the emerging connections makes it clear that software testing for scientific computing is emerging as a distinct domain of interest, with major connections to Cluster 2 and 5. 
    \item As in many other areas of software development, machine learning offers the potential to automate tasks that traditionally require significant human effort, such as test and oracle generation and program repair. 
    \item Test creation practices of many types (including white-box, black-box, mutation, and security) are emerging for web applications.
    \item New test generation approaches are emerging for GUIs and web applications. 
\end{itemize}

\subsection{RQ5: Declining Keywords and Topics}\label{sec:topics_declining}

\begin{figure}[!t]
        \centering
\includegraphics[width=0.82\textwidth]{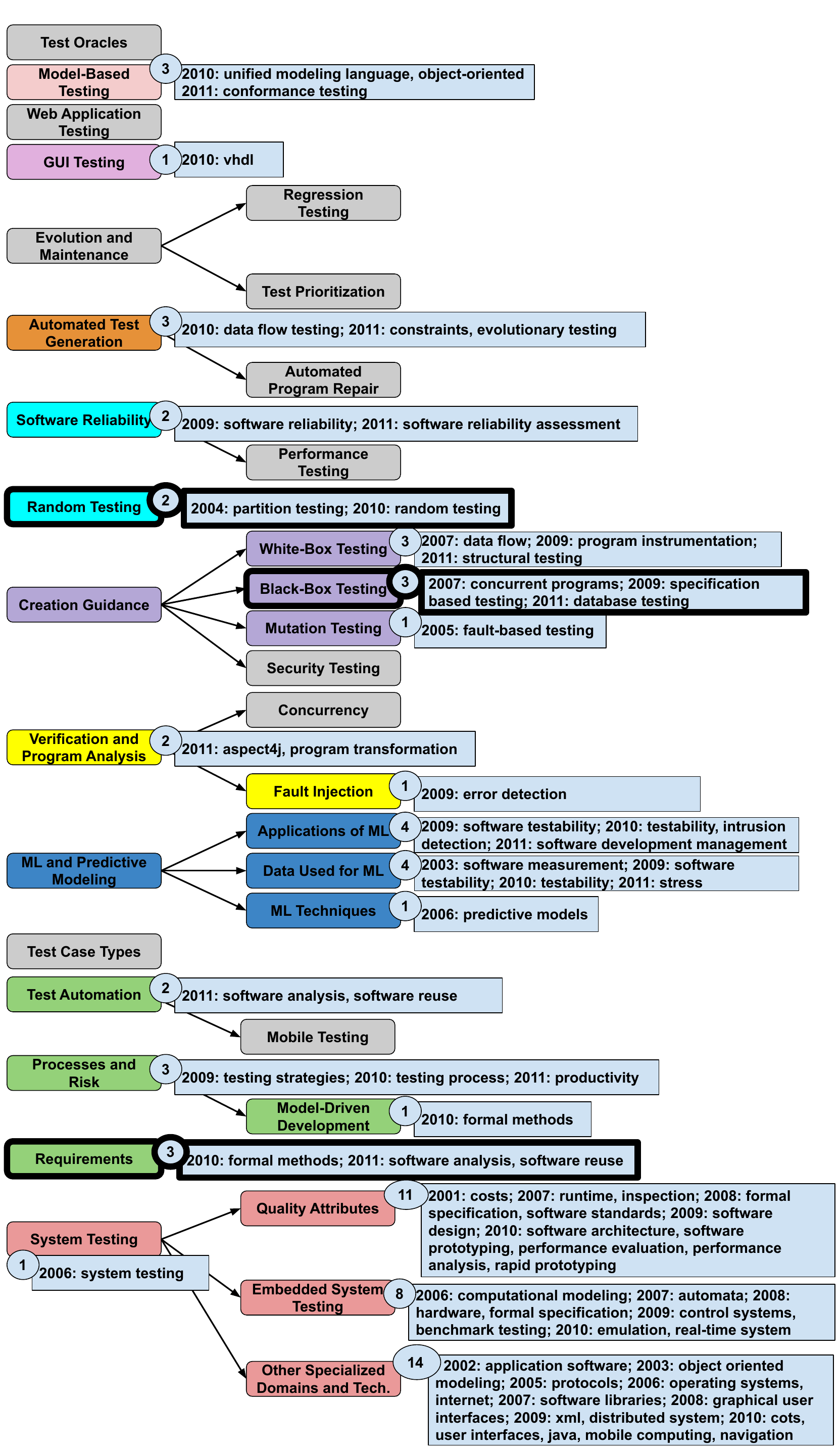}
  \caption{Keywords with an average publication date \textbf{earlier than June 2011}, along with their associated research topic. Topics colored in gray are those without declining keywords. Topics with both declining keywords and a lack of emerging keywords are highlighted.}
    \label{fig:decline_areas} \vspace{-25pt}
\end{figure}

Figure~\ref{fig:decline_areas} shows the 66 keywords with the oldest average date, with their associated research topic. In particular, we highlight three research topics or subtopics that we hypothesize may currently be in decline.

\begin{center}
\begin{framed}
\textbf{RQ5 (Declining Keywords and Topics):} Older average dates of publication and lack of emerging keywords suggest that keywords and topics related to random and requirements-based testing may be in decline.
\end{framed}
\end{center}

\noindent Briefly, we examine these areas:
\begin{itemize}
\setlength{\itemsep}{1pt}
  \setlength{\parskip}{0pt}
  \setlength{\parsep}{0pt} 
    \item Traditional random testing has been supplanted, to some extent, by semi-random approaches. As shown in Figure~\ref{fig:growth_areas}, search-based and fuzzing techniques are growing in popularity. Both use sampling heuristics instead of applying pure random generation, retaining some of the benefits of random testing (e.g., scalability) while potentially yielding more effective results. 
    \item Many of the keywords related to requirements and black-box testing have older average publication dates, indicating potential stagnation. Agile processes favor lightweight requirements (e.g., user stories) over formal and complex requirements. We hypothesize that this may have led to a shift in attention towards other sources of information for test creation.
\end{itemize}

We hesitate to state that these topics are ``dying'' or are solved challenges. However, we do see evidence that they have not seen notable growth in popularity or the emergence of new keywords in recent years. New application areas, techniques, or changes in development processes may lead to a resurgence in interest in the future.

\section{Further Analysis and Advice to Researchers}\label{sec:advice}

Both the high-level topic overview and the low-level map of connections between keywords can serve as inspiration for prospective and experienced researchers. We offer the following advice on how this data could inspire new research.

\smallskip\noindent\textbf{An overview of the testing field:} For inexperienced researchers, the high-level topics offer an immediate ``snapshot'' that can be used to guide exploration of different research areas. The keywords illustrate key concepts that form research topics, and offer targeted suggestions on terms the researcher should examine in detail. Connections between those keywords illustrate how those concepts have been connected in practice, which may encourage critical reflection on both the individual concepts and how they relate. The emerging keywords and topics suggest areas that researchers may wish to pay attention to, and emerging connections clarify how these keywords fit into the field.

\smallskip\noindent\textbf{Understanding the context of a keyword or topic:} Researchers can analyze the map to gain a data-driven view of the field for further planing and development. As a starting point, those interested in a keyword or topic can examine how that keyword or topic fits into the broader context of testing research. 
\begin{itemize}
    \item What keywords are often associated with a keyword of interest? This may illustrate the type of research often conducted on this concept (or its associated topic), and natural areas of synergy between keywords or topics.
    \item Is interest in this keyword or topic growing, declining, or stable? The average date of publication may suggest the current level of interest (or lack thereof).
\end{itemize}

\smallskip\noindent\textbf{Identification of under-explored connections between keywords or topics:} We hypothesize that the map data could potentially inspire future research through analyses of connections between keywords and topics. There are many ways connections could be analyzed. One is to identify keywords that have \textit{under-explored} connections. 

Specifically, an under-explored connection is one where (a) at least one publication has connected the keywords, but (b), the specific number of publications connecting those two keywords is relatively low---indicating potential for additional research exploration. Under-explored connections may serve as inspiration, suggesting concepts that could be connected in further research:
\begin{itemize}
    \item Within a cluster, under-explored connections may suggest ways that concepts within a particular research topic could be more closely linked. For example, different mechanisms from automated test generation algorithms could be blended into hybrid algorithms.\footnote{As has been done for concolic execution and search-based test generation~\cite{Galeotti13:search}.} An examination of under-explored connections could offer similar inspiration.
    \item Across clusters, we could identify either pairs of clusters or topics that could be more deeply connected in future research. In some cases, these may be topics that are already connected (e.g., automated test generation and white-box testing), but where there are opportunities for new research related to specific keywords or aspects of the topic (e.g., specific test generation algorithms).
\end{itemize}
\noindent There are different ways that under-explored connections could be identified and analyzed. As an initial exploration, first, one must identify a lower and upper bound on the number of publications linking keywords. As an example, in the network visualization, four publications are needed for an edge to be shown (by default). Therefore, one could adopt four publications as the threshold for this analysis and capture all connections in a short range of this threshold---e.g., 4-6 publications targeting a pair of keywords. 

\begin{figure}[!t]
        \centering
\includegraphics[width=0.95\textwidth]{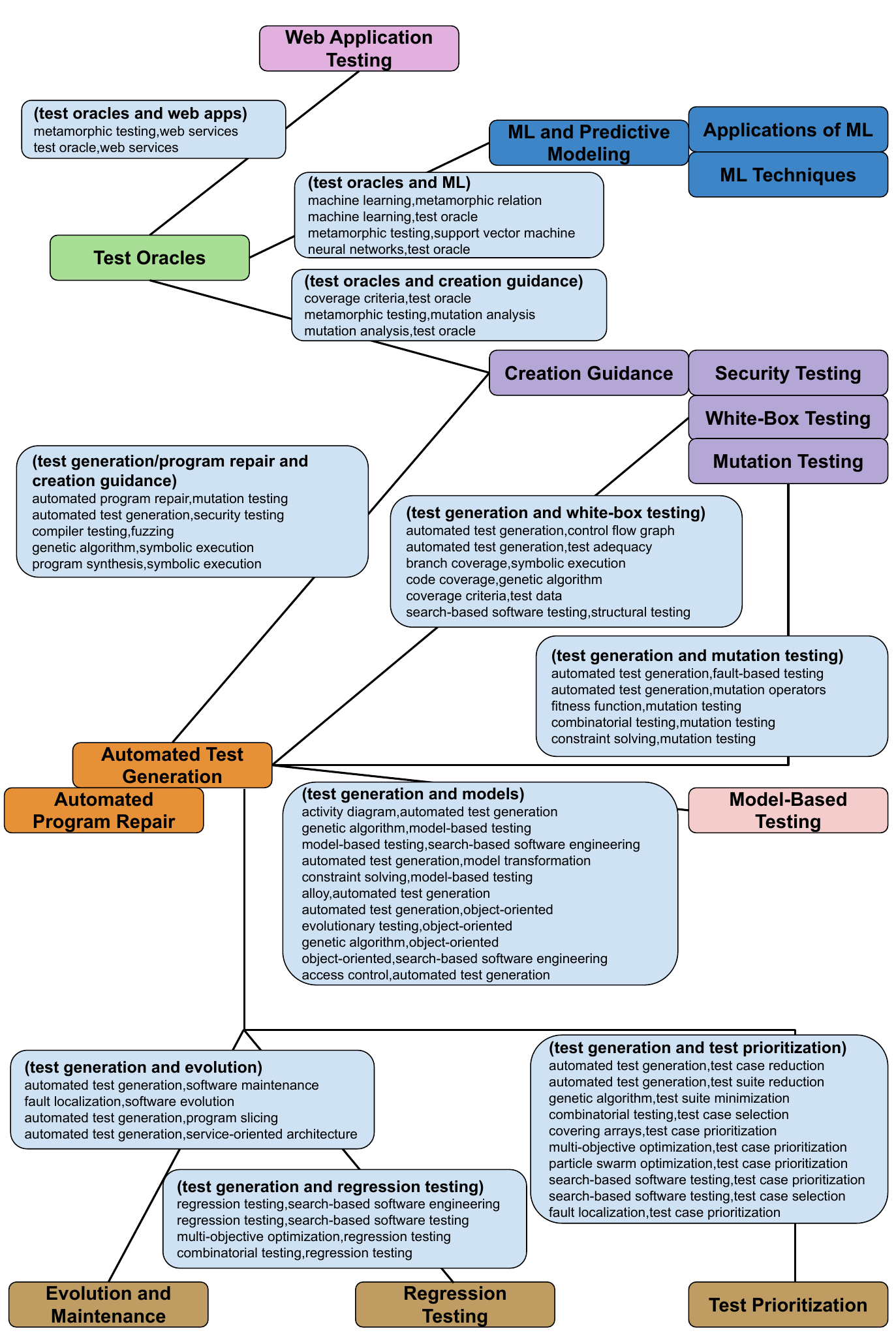} 
  \caption{Under-explored connections (keywords connected by 4-6 publications), connected by research topic, for the six cluster pairings with highest ratio of under-explored to total connections.} 
    \label{fig:underexplored} 
\end{figure}

718 connections have a strength of 4-6 publications. To identify a subset for initial exploration, we can (a) focus on cross-cluster connections, and (b), use the cross-cluster connection density to identify the pairs of clusters with the most under-explored connections. Here, we specifically focus on the cluster pairings where $\geq 2\%$ of all connections between the two clusters consist of under-explored connections. Six cluster pairings met the threshold: Cluster 11 (test oracles) with Cluster 9 (web and GUI testing, 2.63\%), Cluster 5 (creation guidance, 2.34\%), and Cluster 3 (machine learning, 2.08\%), and Cluster 7 (automated test generation) with Cluster 8 (evolution and maintenance, 3.38\%), Cluster 10 (model-based testing, 2.46\%), and Cluster 5 (2.01\%). 

For these cluster pairings, we grouped the connections by research topic, then examined the meaning and potential application of the connections. In Figure~\ref{fig:underexplored}, we illustrate the identified connections, categorized by their associated research topics.

We make several observations about these connections. First, specific suggestions emerge for exploring connections in future research, including (among others):
    \begin{itemize}
        \item The relationship between mutations and test oracles.
        \item Use of mutation as part of automation program repair and test generation.
        \item Test generation based on specific modeling formats (e.g., object-oriented models such as activity diagrams).
        \item Reduction techniques for generated test suites and test cases.
        \item The relationship between test generation and program evolution (e.g., how often tests should be generated, how tests should be maintained). 
        \item Generation of tests for regression testing. 
        \item The use of specific optimization algorithms for test case prioritization. 
    \end{itemize}

In some cases, ``under-explored'' coincides with ``emerging''---for example, test oracles with machine learning and web services. There are also cases where topics are well-connected in research (e.g., test generation and white-box testing) through different keywords (e.g., ``coverage criteria'' instead of ``code coverage''). We retained keywords with minor differences in meaning, as even minor distinctions may be important. However, some connections may be well-explored under a different keyword. Even in such cases, there may be opportunity for further exploration related to these keyword differences, or connections based on concepts and technologies than have not been explored previously (e.g., specific generation algorithms or coverage criteria).

\smallskip\noindent\textbf{Identifying new connections between keywords:} The \textit{absence} of a connection between two keywords does not imply that the concepts cannot be connected. Consider keywords within a single cluster. Keywords lacking a direct connection may represent entirely incompatible concepts. However, in other cases, there may be a natural synergy between the two concepts that had not yet been considered. While the map cannot directly inform researchers which keywords \textit{can} be connected, or how they can be connected, it can serve as a means to prompt brainstorming. 

As an example, we can inspect keywords within a cluster that lack a direct connection to specific other keywords in their cluster. Cluster 8 (evolution and maintenance, with subtopics of regression testing and test prioritization) contains 19 keywords. There are 180 cases where two keywords lack a direct link within Cluster 8---e.g., ``change impact analysis'' and ``test case reduction'' are not directly connected in publications. 

Not all of these cases offer obvious ideas for new research, but consideration of these cases may lead to inspiration. For example, we identified the following ideas:
\begin{itemize}
    \item The use of change impact analysis as part of program comprehension, test case reduction, test suite minimization, or test suite reduction.
    \item The use of information retrieval and natural language processing to provide information for test case and suite reduction, selection, and minimization and for regression test selection. 
    \item The use of regression test selection techniques for use as part of test case and suite reduction and test case selection. 
    \item The use of program comprehension techniques for regression test selection.
    \item The relationship between evolution and maintenance of software with test case prioritization, minimization, and reduction. 
    \item Service-oriented architecture and web services appear in this cluster because of close association with  particular keywords (e.g., regression testing), but are only indirectly connected to the majority of the other keywords. The missing connections suggest the need for targeted test case prioritization, selection, reduction, and minimization approaches for service-oriented architectures and web services, as well as examination of the evolution and maintenance of service-oriented architectures and web services. 
\end{itemize}
\noindent Similar ideas may emerge from inspecting missing connections within other clusters.

There are many ways that this map could potentially be analyzed beyond the simple exploration in this section. We suggest that researchers attempt to analyze different connection types, connection strength thresholds, and other aspects of the collected metadata (e.g., publication age or number of citations) in order to gain inspiration for new research or insight into the field. 


\section{Threats to Validity}\label{sec:threats}

\smallskip\noindent\textbf{Conclusion Validity:} VOSviewer was used to perform visualization. The design of this tool and the visualizations it produces could potentially bias the observations made. 
However, the tool is based on well-understood and established computational principles. Further, it has been used in over 500 bibliometric studies (e.g.,~\cite{Mohammadi20:BigData,Mohammadi12:CoWord}), in a large variety of fields and its assumptions have been verified by experts in these fields.\footnote{A full list of publications is maintained at \url{https://www.vosviewer.com/publications}.} We have made efforts to verify the assumptions behind the analyses performed.

\noindent\textbf{External Validity:} Our study examined publications from the Scopus database, potentially omitting relevant venues for software testing research. Scopus is the one of the most comprehensive databases covering research publications~\cite{Thelwall22:Scopus}, indexing content from 24,600 active conferences or journals and 5,000 publishers.\footnote{List of covered journals and conferences: \url{https://www.scopus.com/sources.uri}.} Specifically, Scopus coverage for computer science research has been found to be better than other databases~\cite{Cavacini15:Scopus}. Scopus also enables efficient export of the data we use to perform our mapping. Although some venues may not be indexed, many of the most important journals and conferences in the software testing field are included. 

We used a single search string to build our sample. Other search strings (e.g., ``software test'') could have complemented the search process. However, our goal is not to capture all studies ever published in software testing. Rather, we require a sufficiently representative sample. We hypothesize that the additional value would be minimal compared to the filtering effort required. We believe that our sample of 57,233 publications is sufficiently large and representative to perform this analysis. 

\smallskip\noindent\textbf{Internal Validity:} We based our analysis on publications retrieved using the term ``software testing''. This pool of papers included publications unrelated to software testing, e.g., the use of software to test hardware or as part of student examination. We performed a manual process to remove unrelated keywords from the mapping. However, it is possible that some publications remain that are unrelated to the targeted research field. We believe that these are not enough to influence our observations.

Our analysis is based on author-supplied keywords, and not other sources of topic information, e.g., titles or abstracts. The use of keywords introduces a risk that publications are mislabeled (e.g., authors used the wrong term), or that important concepts are omitted. Still, author-supplied keywords are a clear and appropriate means to capture the structure of software testing research. Author-supplied keywords are regularly used in other bibliometric analyses~\cite{Mohammadi20:BigData,Jamali20:WineCoWord,Liao18:MedicalCoWord} and have been found to effectively reflect structures in research fields~\cite{Liao18:MedicalCoWord,Lo16:CoWord}. Even if relevant keywords are omitted, the concepts the authors felt were most important are reflected. While there is potential inaccuracy, it is likely that the selected keywords are close to correct. Alternative methods carry similar risks. Automated or external categorization can also be inaccurate and potentially violates the intent of authors. Other sources of information, such as titles or abstracts, introduce noise and are difficult to use to categorize publications. 

We applied a threshold of a minimum of 20 studies before a keyword appeared in our dataset or map. We used this threshold to omit minor or highly obscure keywords and to control the level of noise in the map. This risks also omitting emerging keywords. We tried lower and higher thresholds then we concluded that the current threshold is enough to cover terms with lower frequency and provide a meaningful and lower scatter network of the keywords. It should be noted when we tested lower and higher thresholds, the overall patterns did not change significantly.

\section{Conclusion}\label{sec:conclu}

Testing is the primary means of assessing software correctness and quality. Research in software testing is growing and rapidly-evolving. Based on the keywords assigned to publications, we seek to identify predominant research topics and understand how they are connected and have evolved. 

We have applied co-word analysis to characterize the topology of software testing research over four decades of research publications. In this map, nodes represent keywords, while edges indicate that publications have co-targeted keywords. Nodes are clustered based on density and strength of edges. We examined the most common keywords, summarized clusters into research topics, examined how clusters connect, and identified emerging and declining keywords, topics, and connections. 

We found that the most popular keywords tend to relate to automation, test creation and assessment guidance, assessment of system quality, and cyber-physical systems.
The clusters of keywords suggest that software testing research can be divided into 16 core topics. All topics are connected, but creation guidance, automated test generation, evolution and maintenance, and test oracles have particularly strong connections to other topics, highlighting their multidisciplinary nature. Emerging keywords and topics relate to web and mobile applications, machine learning, energy consumption, automated program repair and test generation, while emerging connections have formed between web applications, test oracles, and machine learning with many topics. Random and requirements-based testing show evidence of decline.

These insights and the underlying map can inspire researchers in software testing by clarifying concepts and their relationships, or by facilitating analyses of the field (e.g., through identification of under-explored and missing connections). In future work, we will broaden the type and scope of analyses of this map data, and we make our data available so that others can do so as well.

\bibliographystyle{elsarticle-num-names}
\bibliography{main}

\begin{thebibliography}{95}
\expandafter\ifx\csname natexlab\endcsname\relax\def\natexlab#1{#1}\fi
\providecommand{\url}[1]{\texttt{#1}}
\providecommand{\href}[2]{#2}
\providecommand{\path}[1]{#1}
\providecommand{\DOIprefix}{doi:}
\providecommand{\ArXivprefix}{arXiv:}
\providecommand{\URLprefix}{URL: }
\providecommand{\Pubmedprefix}{pmid:}
\providecommand{\doi}[1]{\href{http://dx.doi.org/#1}{\path{#1}}}
\providecommand{\Pubmed}[1]{\href{pmid:#1}{\path{#1}}}
\providecommand{\bibinfo}[2]{#2}
\ifx\xfnm\relax \def\xfnm[#1]{\unskip,\space#1}\fi
\bibitem[{Pezze and Young(2006)}]{Pezze06:testing}
\bibinfo{author}{M.~Pezze}, \bibinfo{author}{M.~Young},
  \bibinfo{title}{Software Test and Analysis: Process, Principles, and
  Techniques}, \bibinfo{publisher}{John Wiley and Sons}, \bibinfo{year}{2006}.
\bibitem[{Turing(1989)}]{Turing49:Checking}
\bibinfo{author}{A.~Turing}, \bibinfo{title}{Checking a Large Routine},
  \bibinfo{publisher}{MIT Press}, \bibinfo{address}{Cambridge, MA, USA},
  \bibinfo{year}{1989}, p. \bibinfo{pages}{70–72}.
\bibitem[{Orso and Rothermel(2014)}]{Orso14:STR}
\bibinfo{author}{A.~Orso}, \bibinfo{author}{G.~Rothermel},
\newblock \bibinfo{title}{Software testing: A research travelogue
  (2000--2014)},
\newblock in: \bibinfo{booktitle}{Proceedings of the on Future of Software
  Engineering}, FOSE 2014, \bibinfo{publisher}{ACM}, \bibinfo{address}{New
  York, NY, USA}, \bibinfo{year}{2014}, pp. \bibinfo{pages}{117--132}.
  \URLprefix \url{http://doi.acm.org/10.1145/2593882.2593885}.
  \DOIprefix\doi{10.1145/2593882.2593885}.
\bibitem[{Fortunato et~al.(2018)Fortunato, Bergstrom, B{\"o}rner, Evans,
  Helbing, Milojevi{\'c}, Petersen, Radicchi, Sinatra, Uzzi, Vespignani,
  Waltman, Wang, and Barab{\'a}si}]{Fortunato18:SoS}
\bibinfo{author}{S.~Fortunato}, \bibinfo{author}{C.~T. Bergstrom},
  \bibinfo{author}{K.~B{\"o}rner}, \bibinfo{author}{J.~A. Evans},
  \bibinfo{author}{D.~Helbing}, \bibinfo{author}{S.~Milojevi{\'c}},
  \bibinfo{author}{A.~M. Petersen}, \bibinfo{author}{F.~Radicchi},
  \bibinfo{author}{R.~Sinatra}, \bibinfo{author}{B.~Uzzi},
  \bibinfo{author}{A.~Vespignani}, \bibinfo{author}{L.~Waltman},
  \bibinfo{author}{D.~Wang}, \bibinfo{author}{A.-L. Barab{\'a}si},
\newblock \bibinfo{title}{Science of science},
\newblock \bibinfo{journal}{Science} \bibinfo{volume}{359}
  (\bibinfo{year}{2018}). \URLprefix
  \url{https://science.sciencemag.org/content/359/6379/eaao0185}.
  \DOIprefix\doi{10.1126/science.aao0185}.
  \href{http://arxiv.org/abs/https://science.sciencemag.org/content/359/6379/eaao0185.full.pdf}{{\tt
  arXiv:https://science.sciencemag.org/content/359/6379/eaao0185.full.pdf}}.
\bibitem[{Borgman and Furner(2002)}]{Borgman02:Biblio}
\bibinfo{author}{C.~L. Borgman}, \bibinfo{author}{J.~Furner},
\newblock \bibinfo{title}{Scholarly communication and bibliometrics},
\newblock \bibinfo{journal}{Annual Review of Information Science and
  Technology} \bibinfo{volume}{36} (\bibinfo{year}{2002})
  \bibinfo{pages}{2--72}. \URLprefix
  \url{https://asistdl.onlinelibrary.wiley.com/doi/abs/10.1002/aris.1440360102}.
  \DOIprefix\doi{https://doi.org/10.1002/aris.1440360102}.
  \href{http://arxiv.org/abs/https://asistdl.onlinelibrary.wiley.com/doi/pdf/10.1002/aris.1440360102}{{\tt
  arXiv:https://asistdl.onlinelibrary.wiley.com/doi/pdf/10.1002/aris.1440360102}}.
\bibitem[{Moed(2006)}]{Moed06:Citation}
\bibinfo{author}{H.~F. Moed}, \bibinfo{title}{Citation analysis in research
  evaluation}, volume~\bibinfo{volume}{9}, \bibinfo{publisher}{Springer Science
  \& Business Media}, \bibinfo{year}{2006}.
\bibitem[{Ding et~al.(2013)Ding, Liu, Guo, and Cronin}]{Ding13:References}
\bibinfo{author}{Y.~Ding}, \bibinfo{author}{X.~Liu}, \bibinfo{author}{C.~Guo},
  \bibinfo{author}{B.~Cronin},
\newblock \bibinfo{title}{The distribution of references across texts: Some
  implications for citation analysis},
\newblock \bibinfo{journal}{Journal of Informetrics} \bibinfo{volume}{7}
  (\bibinfo{year}{2013}) \bibinfo{pages}{583--592}. \URLprefix
  \url{https://www.sciencedirect.com/science/article/pii/S1751157713000230}.
  \DOIprefix\doi{https://doi.org/10.1016/j.joi.2013.03.003}.
\bibitem[{Pritchard et~al.(1969)}]{pritchard1969statistical}
\bibinfo{author}{A.~Pritchard}, et~al.,
\newblock \bibinfo{title}{Statistical bibliography or bibliometrics},
\newblock \bibinfo{journal}{Journal of documentation} \bibinfo{volume}{25}
  (\bibinfo{year}{1969}) \bibinfo{pages}{348--349}.
\bibitem[{De~Bellis(2009)}]{de2009bibliometrics}
\bibinfo{author}{N.~De~Bellis}, \bibinfo{title}{Bibliometrics and citation
  analysis: from the science citation index to cybermetrics},
  \bibinfo{publisher}{scarecrow press}, \bibinfo{year}{2009}.
\bibitem[{Donthu et~al.(2020)Donthu, Kumar, and Pattnaik}]{donthu2020forty}
\bibinfo{author}{N.~Donthu}, \bibinfo{author}{S.~Kumar},
  \bibinfo{author}{D.~Pattnaik},
\newblock \bibinfo{title}{Forty-five years of journal of business research: a
  bibliometric analysis},
\newblock \bibinfo{journal}{Journal of Business Research} \bibinfo{volume}{109}
  (\bibinfo{year}{2020}) \bibinfo{pages}{1--14}.
\bibitem[{Van~Eck and Waltman(2014)}]{van2014visualizing}
\bibinfo{author}{N.~J. Van~Eck}, \bibinfo{author}{L.~Waltman},
\newblock \bibinfo{title}{Visualizing bibliometric networks},
\newblock in: \bibinfo{booktitle}{Measuring scholarly impact},
  \bibinfo{publisher}{Springer}, \bibinfo{year}{2014}, pp.
  \bibinfo{pages}{285--320}.
\bibitem[{Peters and {van Raan}(1993)}]{Peters93:CoWord}
\bibinfo{author}{H.~Peters}, \bibinfo{author}{A.~{van Raan}},
\newblock \bibinfo{title}{Co-word-based science maps of chemical engineering.
  part i: Representations by direct multidimensional scaling},
\newblock \bibinfo{journal}{Research Policy} \bibinfo{volume}{22}
  (\bibinfo{year}{1993}) \bibinfo{pages}{23--45}. \URLprefix
  \url{https://www.sciencedirect.com/science/article/pii/004873339390031C}.
  \DOIprefix\doi{https://doi.org/10.1016/0048-7333(93)90031-C}.
\bibitem[{Whittaker(1989)}]{whittaker1989history}
\bibinfo{author}{E.~Whittaker}, \bibinfo{title}{A History of the Theories of
  Aether and Electricity: Vol. I: The Classical Theories; Vol. II: The Modern
  Theories, 1900-1926}, volume~\bibinfo{volume}{1}, \bibinfo{publisher}{Courier
  Dover Publications}, \bibinfo{year}{1989}.
\bibitem[{Su and Lee(2010)}]{su2010mapping}
\bibinfo{author}{H.-N. Su}, \bibinfo{author}{P.-C. Lee},
\newblock \bibinfo{title}{Mapping knowledge structure by keyword co-occurrence:
  a first look at journal papers in technology foresight},
\newblock \bibinfo{journal}{Scientometrics} \bibinfo{volume}{85}
  (\bibinfo{year}{2010}) \bibinfo{pages}{65--79}.
\bibitem[{Romo-Fern{\'a}ndez et~al.(2013)Romo-Fern{\'a}ndez, Guerrero-Bote, and
  Moya-Aneg{\'o}n}]{Romo13:co}
\bibinfo{author}{L.~M. Romo-Fern{\'a}ndez}, \bibinfo{author}{V.~P.
  Guerrero-Bote}, \bibinfo{author}{F.~Moya-Aneg{\'o}n},
\newblock \bibinfo{title}{Co-word based thematic analysis of renewable energy
  (1990--2010)},
\newblock \bibinfo{journal}{Scientometrics} \bibinfo{volume}{97}
  (\bibinfo{year}{2013}) \bibinfo{pages}{743--765}.
\bibitem[{Marx et~al.(2017)Marx, Haunschild, and Bornmann}]{Werner17:CoWord}
\bibinfo{author}{W.~Marx}, \bibinfo{author}{R.~Haunschild},
  \bibinfo{author}{L.~Bornmann},
\newblock \bibinfo{title}{Global warming and tea production—the bibliometric
  view on a newly emerging research topic},
\newblock \bibinfo{journal}{Climate} \bibinfo{volume}{5}
  (\bibinfo{year}{2017}). \URLprefix
  \url{https://www.mdpi.com/2225-1154/5/3/46}.
  \DOIprefix\doi{10.3390/cli5030046}.
\bibitem[{Mohammadi(2012)}]{Mohammadi12:CoWord}
\bibinfo{author}{E.~Mohammadi},
\newblock \bibinfo{title}{Knowledge mapping of the iranian nanoscience and
  technology: a text mining approach},
\newblock \bibinfo{journal}{Scientometrics Scientometrics} \bibinfo{volume}{92}
  (\bibinfo{year}{01 Sep. 2012}) \bibinfo{pages}{593 -- 608}. \URLprefix
  \url{https://akjournals.com/view/journals/11192/92/3/article-p593.xml}.
  \DOIprefix\doi{10.1007/s11192-012-0644-6}.
\bibitem[{Liu et~al.(2014)Liu, Goncalves, Ferreira, Xiao, Hosio, and
  Kostakos}]{Liu14:CHICoWord}
\bibinfo{author}{Y.~Liu}, \bibinfo{author}{J.~Goncalves},
  \bibinfo{author}{D.~Ferreira}, \bibinfo{author}{B.~Xiao},
  \bibinfo{author}{S.~Hosio}, \bibinfo{author}{V.~Kostakos},
\newblock \bibinfo{title}{Chi 1994-2013: Mapping two decades of intellectual
  progress through co-word analysis},
\newblock in: \bibinfo{booktitle}{Proceedings of the SIGCHI Conference on Human
  Factors in Computing Systems}, CHI '14, \bibinfo{publisher}{Association for
  Computing Machinery}, \bibinfo{address}{New York, NY, USA},
  \bibinfo{year}{2014}, p. \bibinfo{pages}{3553–3562}. \URLprefix
  \url{https://doi.org/10.1145/2556288.2556969}.
  \DOIprefix\doi{10.1145/2556288.2556969}.
\bibitem[{Liao et~al.(2018)Liao, Tang, Luo, Li, Chiclana, and
  Zeng}]{Liao18:MedicalCoWord}
\bibinfo{author}{H.~Liao}, \bibinfo{author}{M.~Tang}, \bibinfo{author}{L.~Luo},
  \bibinfo{author}{C.~Li}, \bibinfo{author}{F.~Chiclana},
  \bibinfo{author}{X.-J. Zeng},
\newblock \bibinfo{title}{A bibliometric analysis and visualization of medical
  big data research},
\newblock \bibinfo{journal}{Sustainability} \bibinfo{volume}{10}
  (\bibinfo{year}{2018}). \URLprefix
  \url{https://www.mdpi.com/2071-1050/10/1/166}.
  \DOIprefix\doi{10.3390/su10010166}.
\bibitem[{Mohammadi and Karami(2020)}]{Mohammadi20:BigData}
\bibinfo{author}{E.~Mohammadi}, \bibinfo{author}{A.~Karami},
\newblock \bibinfo{title}{Exploring research trends in big data across
  disciplines: A text mining analysis},
\newblock \bibinfo{journal}{Journal of Information Science} \bibinfo{volume}{0}
  (\bibinfo{year}{2020}) \bibinfo{pages}{0165551520932855}. \URLprefix
  \url{https://doi.org/10.1177/0165551520932855}.
  \DOIprefix\doi{10.1177/0165551520932855}.
  \href{http://arxiv.org/abs/https://doi.org/10.1177/0165551520932855}{{\tt
  arXiv:https://doi.org/10.1177/0165551520932855}}.
\bibitem[{Garousi and Mäntylä(2016)}]{Garousi16:SEBib}
\bibinfo{author}{V.~Garousi}, \bibinfo{author}{M.~V. Mäntylä},
\newblock \bibinfo{title}{Citations, research topics and active countries in
  software engineering: A bibliometrics study},
\newblock \bibinfo{journal}{Computer Science Review} \bibinfo{volume}{19}
  (\bibinfo{year}{2016}) \bibinfo{pages}{56--77}. \URLprefix
  \url{https://www.sciencedirect.com/science/article/pii/S1574013715300654}.
  \DOIprefix\doi{https://doi.org/10.1016/j.cosrev.2015.12.002}.
\bibitem[{Garousi and Fernandes(2017)}]{Garousi17:Quantity}
\bibinfo{author}{V.~Garousi}, \bibinfo{author}{J.~M. Fernandes},
\newblock \bibinfo{title}{Quantity versus impact of software engineering
  papers: a quantitative study},
\newblock \bibinfo{journal}{Scientometrics} \bibinfo{volume}{112}
  (\bibinfo{year}{2017}) \bibinfo{pages}{963--1006}. \URLprefix
  \url{https://doi.org/10.1007/s11192-017-2419-6}.
  \DOIprefix\doi{10.1007/s11192-017-2419-6}.
\bibitem[{Karanatsiou et~al.(2019)Karanatsiou, Li, Arvanitou, Misirlis, and
  Wong}]{Karanatsiou19:SE}
\bibinfo{author}{D.~Karanatsiou}, \bibinfo{author}{Y.~Li},
  \bibinfo{author}{E.-M. Arvanitou}, \bibinfo{author}{N.~Misirlis},
  \bibinfo{author}{W.~E. Wong},
\newblock \bibinfo{title}{A bibliometric assessment of software engineering
  scholars and institutions (2010–2017)},
\newblock \bibinfo{journal}{Journal of Systems and Software}
  \bibinfo{volume}{147} (\bibinfo{year}{2019}) \bibinfo{pages}{246--261}.
  \URLprefix
  \url{https://www.sciencedirect.com/science/article/pii/S0164121218302334}.
  \DOIprefix\doi{https://doi.org/10.1016/j.jss.2018.10.029}.
\bibitem[{Wong et~al.(2008)Wong, Tse, Glass, Basili, and
  Chen}]{Wong08:Scholars}
\bibinfo{author}{W.~E. Wong}, \bibinfo{author}{T.~Tse}, \bibinfo{author}{R.~L.
  Glass}, \bibinfo{author}{V.~R. Basili}, \bibinfo{author}{T.~Chen},
\newblock \bibinfo{title}{An assessment of systems and software engineering
  scholars and institutions (2001–2005)},
\newblock \bibinfo{journal}{Journal of Systems and Software}
  \bibinfo{volume}{81} (\bibinfo{year}{2008}) \bibinfo{pages}{1059--1062}.
  \URLprefix
  \url{https://www.sciencedirect.com/science/article/pii/S0164121207002300}.
  \DOIprefix\doi{https://doi.org/10.1016/j.jss.2007.09.018},
  \bibinfo{note}{agile Product Line Engineering}.
\bibitem[{Wong et~al.(2009)Wong, Tse, Glass, Basili, and
  Chen}]{Wong09:Scholars}
\bibinfo{author}{W.~E. Wong}, \bibinfo{author}{T.~Tse}, \bibinfo{author}{R.~L.
  Glass}, \bibinfo{author}{V.~R. Basili}, \bibinfo{author}{T.~Chen},
\newblock \bibinfo{title}{An assessment of systems and software engineering
  scholars and institutions (2002–2006)},
\newblock \bibinfo{journal}{Journal of Systems and Software}
  \bibinfo{volume}{82} (\bibinfo{year}{2009}) \bibinfo{pages}{1370--1373}.
  \URLprefix
  \url{https://www.sciencedirect.com/science/article/pii/S0164121209001265}.
  \DOIprefix\doi{https://doi.org/10.1016/j.jss.2009.06.018}, \bibinfo{note}{sI:
  Architectural Decisions and Rationale}.
\bibitem[{Wong et~al.(2011)Wong, Tse, Glass, Basili, and
  Chen}]{Wong11:Scholars}
\bibinfo{author}{W.~E. Wong}, \bibinfo{author}{T.~Tse}, \bibinfo{author}{R.~L.
  Glass}, \bibinfo{author}{V.~R. Basili}, \bibinfo{author}{T.~Chen},
\newblock \bibinfo{title}{An assessment of systems and software engineering
  scholars and institutions (2003–2007 and 2004–2008)},
\newblock \bibinfo{journal}{Journal of Systems and Software}
  \bibinfo{volume}{84} (\bibinfo{year}{2011}) \bibinfo{pages}{162--168}.
  \URLprefix
  \url{https://www.sciencedirect.com/science/article/pii/S0164121210002682}.
  \DOIprefix\doi{https://doi.org/10.1016/j.jss.2010.09.036},
  \bibinfo{note}{information Networking and Software Services}.
\bibitem[{Garousi and Varma(2010)}]{Garousi10:Canada}
\bibinfo{author}{V.~Garousi}, \bibinfo{author}{T.~Varma},
\newblock \bibinfo{title}{A bibliometric assessment of canadian software
  engineering scholars and institutions (1996-2006)},
\newblock \bibinfo{journal}{Computer and Information Science}
  \bibinfo{volume}{3} (\bibinfo{year}{2010}) \bibinfo{pages}{19}.
\bibitem[{Garousi(2015)}]{Garousi15:SETurkish}
\bibinfo{author}{V.~Garousi},
\newblock \bibinfo{title}{A bibliometric analysis of the turkish software
  engineering research community},
\newblock \bibinfo{journal}{Scientometrics} \bibinfo{volume}{105}
  (\bibinfo{year}{2015}) \bibinfo{pages}{23--49}. \URLprefix
  \url{https://doi.org/10.1007/s11192-015-1663-x}.
  \DOIprefix\doi{10.1007/s11192-015-1663-x}.
\bibitem[{Farhoodi et~al.(2013)Farhoodi, Garousi, Pfahl, and
  Sillito}]{Farhoodi13:SciSoft}
\bibinfo{author}{R.~Farhoodi}, \bibinfo{author}{V.~Garousi},
  \bibinfo{author}{D.~Pfahl}, \bibinfo{author}{J.~Sillito},
\newblock \bibinfo{title}{Development of scientific software: A systematic
  mapping, a bibliometrics study, and a paper repository},
\newblock \bibinfo{journal}{International Journal of Software Engineering and
  Knowledge Engineering} \bibinfo{volume}{23} (\bibinfo{year}{2013})
  \bibinfo{pages}{463--506}. \URLprefix
  \url{https://doi.org/10.1142/S0218194013500137}.
  \DOIprefix\doi{10.1142/S0218194013500137}.
  \href{http://arxiv.org/abs/https://doi.org/10.1142/S0218194013500137}{{\tt
  arXiv:https://doi.org/10.1142/S0218194013500137}}.
\bibitem[{de~Freitas and de~Souza(2011)}]{deFreitas11:SBSE}
\bibinfo{author}{F.~G. de~Freitas}, \bibinfo{author}{J.~T. de~Souza},
\newblock \bibinfo{title}{Ten years of search based software engineering: A
  bibliometric analysis},
\newblock in: \bibinfo{editor}{M.~B. Cohen},
  \bibinfo{editor}{M.~{\'O}~Cinn{\'e}ide} (Eds.), \bibinfo{booktitle}{Search
  Based Software Engineering}, \bibinfo{publisher}{Springer Berlin Heidelberg},
  \bibinfo{address}{Berlin, Heidelberg}, \bibinfo{year}{2011}, pp.
  \bibinfo{pages}{18--32}.
\bibitem[{Harrold(2000)}]{Harrold00:roadmap}
\bibinfo{author}{M.~J. Harrold},
\newblock \bibinfo{title}{Testing: A roadmap},
\newblock in: \bibinfo{booktitle}{Proceedings of the Conference on The Future
  of Software Engineering}, ICSE '00, \bibinfo{publisher}{Association for
  Computing Machinery}, \bibinfo{address}{New York, NY, USA},
  \bibinfo{year}{2000}, p. \bibinfo{pages}{61–72}. \URLprefix
  \url{https://doi.org/10.1145/336512.336532}.
  \DOIprefix\doi{10.1145/336512.336532}.
\bibitem[{{Bertolino}(2007)}]{Bertolino07:Achievements}
\bibinfo{author}{A.~{Bertolino}},
\newblock \bibinfo{title}{Software testing research: Achievements, challenges,
  dreams},
\newblock in: \bibinfo{booktitle}{Future of Software Engineering (FOSE '07)},
  \bibinfo{year}{2007}, pp. \bibinfo{pages}{85--103}.
  \DOIprefix\doi{10.1109/FOSE.2007.25}.
\bibitem[{Van~Eck and Waltman(2010)}]{Van10:VOSviewer}
\bibinfo{author}{N.~J. Van~Eck}, \bibinfo{author}{L.~Waltman},
\newblock \bibinfo{title}{Software survey: Vosviewer, a computer program for
  bibliometric mapping},
\newblock \bibinfo{journal}{scientometrics} \bibinfo{volume}{84}
  (\bibinfo{year}{2010}) \bibinfo{pages}{523--538}.
\bibitem[{van Eck and Waltman(2014)}]{VanEck14:Visualizing}
\bibinfo{author}{N.~J. van Eck}, \bibinfo{author}{L.~Waltman},
  \bibinfo{title}{Visualizing Bibliometric Networks},
  \bibinfo{publisher}{Springer International Publishing},
  \bibinfo{address}{Cham}, \bibinfo{year}{2014}, pp. \bibinfo{pages}{285--320}.
  \URLprefix \url{https://doi.org/10.1007/978-3-319-10377-8_13}.
  \DOIprefix\doi{10.1007/978-3-319-10377-8_13}.
\bibitem[{Anand et~al.(2013)Anand, Burke, Chen, Clark, Cohen, Grieskamp,
  Harman, Harrold, and McMinn}]{Anand13:Orchestrated}
\bibinfo{author}{S.~Anand}, \bibinfo{author}{E.~K. Burke},
  \bibinfo{author}{T.~Y. Chen}, \bibinfo{author}{J.~Clark},
  \bibinfo{author}{M.~B. Cohen}, \bibinfo{author}{W.~Grieskamp},
  \bibinfo{author}{M.~Harman}, \bibinfo{author}{M.~J. Harrold},
  \bibinfo{author}{P.~McMinn},
\newblock \bibinfo{title}{An orchestrated survey of methodologies for automated
  software test case generation},
\newblock \bibinfo{journal}{Journal of Systems and Software}
  \bibinfo{volume}{86} (\bibinfo{year}{2013}) \bibinfo{pages}{1978--2001}.
\bibitem[{Korel and Al-Yami(1998)}]{Korel98:ART}
\bibinfo{author}{B.~Korel}, \bibinfo{author}{A.~M. Al-Yami},
\newblock \bibinfo{title}{Automated regression test generation},
\newblock in: \bibinfo{booktitle}{Proceedings of the 1998 ACM SIGSOFT
  International Symposium on Software Testing and Analysis}, ISSTA '98,
  \bibinfo{publisher}{ACM}, \bibinfo{address}{New York, NY, USA},
  \bibinfo{year}{1998}, pp. \bibinfo{pages}{143--152}. \URLprefix
  \url{http://doi.acm.org/10.1145/271771.271803}.
  \DOIprefix\doi{10.1145/271771.271803}.
\bibitem[{Just(2014)}]{Just14:MMF}
\bibinfo{author}{R.~Just},
\newblock \bibinfo{title}{The major mutation framework: Efficient and scalable
  mutation analysis for java},
\newblock in: \bibinfo{booktitle}{Proceedings of the 2014 International
  Symposium on Software Testing and Analysis}, ISSTA 2014,
  \bibinfo{publisher}{ACM}, \bibinfo{address}{New York, NY, USA},
  \bibinfo{year}{2014}, pp. \bibinfo{pages}{433--436}. \URLprefix
  \url{http://doi.acm.org/10.1145/2610384.2628053}.
  \DOIprefix\doi{10.1145/2610384.2628053}.
\bibitem[{Fewster and Graham(1999)}]{Fewster99:Automation}
\bibinfo{author}{M.~Fewster}, \bibinfo{author}{D.~Graham},
  \bibinfo{title}{Software test automation}, \bibinfo{publisher}{Addison-Wesley
  Reading}, \bibinfo{year}{1999}.
\bibitem[{Mitchell(1998)}]{Mitchell98:IGA}
\bibinfo{author}{M.~Mitchell}, \bibinfo{title}{An Introduction to Genetic
  Algorithms}, \bibinfo{publisher}{MIT Press}, \bibinfo{address}{Cambridge, MA,
  USA}, \bibinfo{year}{1998}.
\bibitem[{Voas(1997)}]{Voas97:FaultInjection}
\bibinfo{author}{J.~Voas},
\newblock \bibinfo{title}{Fault injection for the masses},
\newblock \bibinfo{journal}{Computer} \bibinfo{volume}{30}
  (\bibinfo{year}{1997}) \bibinfo{pages}{129--130}.
  \DOIprefix\doi{10.1109/2.642820}.
\bibitem[{Kitchenham and Pfleeger(1996)}]{Kitchenham96:Quality}
\bibinfo{author}{B.~Kitchenham}, \bibinfo{author}{S.~Pfleeger},
\newblock \bibinfo{title}{Software quality: the elusive target [special issues
  section]},
\newblock \bibinfo{journal}{IEEE Software} \bibinfo{volume}{13}
  (\bibinfo{year}{1996}) \bibinfo{pages}{12--21}.
  \DOIprefix\doi{10.1109/52.476281}.
\bibitem[{Herzner et~al.(2007)Herzner, Schlick, Le~Guennec, and
  Martin}]{Herzner05}
\bibinfo{author}{W.~Herzner}, \bibinfo{author}{R.~Schlick},
  \bibinfo{author}{A.~Le~Guennec}, \bibinfo{author}{B.~Martin},
\newblock \bibinfo{title}{Model-based simulation of distributed real-time
  applications},
\newblock in: \bibinfo{booktitle}{5th IEEE Int'l Conf. on Industrial
  Infomatics}, \bibinfo{year}{2007}, pp. \bibinfo{pages}{989 -- 994}.
\bibitem[{Mok and Stuart(1996)}]{Mok96:SimulationVerification}
\bibinfo{author}{A.~K. Mok}, \bibinfo{author}{D.~Stuart},
\newblock \bibinfo{title}{Simulation vs. verification: Getting the best of both
  worlds},
\newblock in: \bibinfo{booktitle}{Proceedings of the Eleventh Annual Conf. on
  Computer Assurance, COMPASS 96}, \bibinfo{year}{1996}.
\bibitem[{Gay et~al.(2017)Gay, Rayadurgam, and Heimdahl}]{Gay16:SteeringTSE}
\bibinfo{author}{G.~Gay}, \bibinfo{author}{S.~Rayadurgam},
  \bibinfo{author}{M.~P.~E. Heimdahl},
\newblock \bibinfo{title}{Automated steering of model-based test oracles to
  admit real program behaviors},
\newblock \bibinfo{journal}{IEEE Transactions on Software Engineering}
  \bibinfo{volume}{43} (\bibinfo{year}{2017}) \bibinfo{pages}{531--555}.
  \DOIprefix\doi{10.1109/TSE.2016.2615311}.
\bibitem[{Crossley(2000)}]{Crossley00:Quality}
\bibinfo{author}{M.~Crossley}, \bibinfo{title}{The Desk Reference of
  Statistical Quality Methods}, \bibinfo{publisher}{ASQ Quality Press},
  \bibinfo{year}{2000}.
\bibitem[{Rothermel et~al.(2002)Rothermel, Harrold, von Ronne, and
  Hong}]{Rothermel02:Reduction}
\bibinfo{author}{G.~Rothermel}, \bibinfo{author}{M.~J. Harrold},
  \bibinfo{author}{J.~von Ronne}, \bibinfo{author}{C.~Hong},
\newblock \bibinfo{title}{Empirical studies of test-suite reduction},
\newblock \bibinfo{journal}{Software Testing, Verification and Reliability}
  \bibinfo{volume}{12} (\bibinfo{year}{2002}) \bibinfo{pages}{219--249}.
  \URLprefix \url{http://dx.doi.org/10.1002/stvr.256}.
  \DOIprefix\doi{10.1002/stvr.256}.
\bibitem[{Gay et~al.(2015)Gay, Staats, Whalen, and Heimdahl}]{Gay15:risks}
\bibinfo{author}{G.~Gay}, \bibinfo{author}{M.~Staats},
  \bibinfo{author}{M.~Whalen}, \bibinfo{author}{M.~Heimdahl},
\newblock \bibinfo{title}{The risks of coverage-directed test case generation},
\newblock \bibinfo{journal}{Software Engineering, IEEE Transactions on}
  \bibinfo{volume}{PP} (\bibinfo{year}{2015}).
  \DOIprefix\doi{10.1109/TSE.2015.2421011}.
\bibitem[{Nie and Leung(2011)}]{Nie11:Combinatorial}
\bibinfo{author}{C.~Nie}, \bibinfo{author}{H.~Leung},
\newblock \bibinfo{title}{A survey of combinatorial testing},
\newblock \bibinfo{journal}{ACM Comput. Surv.} \bibinfo{volume}{43}
  (\bibinfo{year}{2011}). \URLprefix
  \url{https://doi.org/10.1145/1883612.1883618}.
  \DOIprefix\doi{10.1145/1883612.1883618}.
\bibitem[{Barr et~al.(2015)Barr, Harman, McMinn, Shahbaz, and
  Yoo}]{Harman13:oraclesurvey}
\bibinfo{author}{E.~Barr}, \bibinfo{author}{M.~Harman},
  \bibinfo{author}{P.~McMinn}, \bibinfo{author}{M.~Shahbaz},
  \bibinfo{author}{S.~Yoo},
\newblock \bibinfo{title}{The oracle problem in software testing: A survey},
\newblock \bibinfo{journal}{IEEE Transactions on Software Engineering}
  \bibinfo{volume}{41} (\bibinfo{year}{2015}) \bibinfo{pages}{507--525}.
  \DOIprefix\doi{10.1109/TSE.2014.2372785}.
\bibitem[{Turhan et~al.(2009)Turhan, Menzies, Bener, and
  Di~Stefano}]{Turhan09:CCWC}
\bibinfo{author}{B.~Turhan}, \bibinfo{author}{T.~Menzies},
  \bibinfo{author}{A.~B. Bener}, \bibinfo{author}{J.~Di~Stefano},
\newblock \bibinfo{title}{On the relative value of cross-company and
  within-company data for defect prediction},
\newblock \bibinfo{journal}{Empirical Software Engineering}
  \bibinfo{volume}{14} (\bibinfo{year}{2009}) \bibinfo{pages}{540--578}.
\bibitem[{Salahirad et~al.(2019)Salahirad, Almulla, and Gay}]{Gay19:fitness}
\bibinfo{author}{A.~Salahirad}, \bibinfo{author}{H.~Almulla},
  \bibinfo{author}{G.~Gay},
\newblock \bibinfo{title}{Choosing the fitness function for the job: Automated
  generation of test suites that detect real faults},
\newblock \bibinfo{journal}{Software Testing, Verification and Reliability}
  \bibinfo{volume}{29} (\bibinfo{year}{2019}) \bibinfo{pages}{e1701}.
  \URLprefix \url{https://onlinelibrary.wiley.com/doi/abs/10.1002/stvr.1701}.
  \DOIprefix\doi{10.1002/stvr.1701}.
  \href{http://arxiv.org/abs/https://onlinelibrary.wiley.com/doi/pdf/10.1002/stvr.1701}{{\tt
  arXiv:https://onlinelibrary.wiley.com/doi/pdf/10.1002/stvr.1701}},
  \bibinfo{note}{e1701 stvr.1701}.
\bibitem[{Cadar et~al.(2008)Cadar, Dunbar, and Engler}]{Cadar:2008:KLEE}
\bibinfo{author}{C.~Cadar}, \bibinfo{author}{D.~Dunbar},
  \bibinfo{author}{D.~Engler},
\newblock \bibinfo{title}{Klee: Unassisted and automatic generation of
  high-coverage tests for complex systems programs},
\newblock in: \bibinfo{booktitle}{Proceedings of the 8th USENIX Conference on
  Operating Systems Design and Implementation}, OSDI'08,
  \bibinfo{publisher}{USENIX Association}, \bibinfo{address}{Berkeley, CA,
  USA}, \bibinfo{year}{2008}, pp. \bibinfo{pages}{209--224}. \URLprefix
  \url{http://dl.acm.org/citation.cfm?id=1855741.1855756}.
\bibitem[{Zander et~al.(2011)Zander, Schieferdecker, and
  Mosterman}]{Zander11:MBT}
\bibinfo{author}{J.~Zander}, \bibinfo{author}{I.~Schieferdecker},
  \bibinfo{author}{P.~J. Mosterman}, \bibinfo{title}{Model-based testing for
  embedded systems}, \bibinfo{publisher}{CRC press}, \bibinfo{year}{2011}.
\bibitem[{Fukuda(1992)}]{Fukuda92}
\bibinfo{author}{T.~Fukuda},
\newblock \bibinfo{title}{Theory and applications of neural networks for
  industrial control systems},
\newblock \bibinfo{journal}{IEEE Transactions on Industrial Electronics}
  (\bibinfo{year}{1992}) \bibinfo{pages}{472--489}.
\bibitem[{van Lamsweerde(2007)}]{vanLamsweerde07:kaos}
\bibinfo{author}{A.~van Lamsweerde},
\newblock \bibinfo{title}{Engineering requirements for system reliability and
  security},
\newblock \bibinfo{journal}{Software System Reliability and Security}
  \bibinfo{volume}{9} (\bibinfo{year}{2007}).
\bibitem[{Jamil et~al.(2016)Jamil, Arif, Abubakar, and
  Ahmad}]{jamil2016software}
\bibinfo{author}{M.~A. Jamil}, \bibinfo{author}{M.~Arif},
  \bibinfo{author}{N.~S.~A. Abubakar}, \bibinfo{author}{A.~Ahmad},
\newblock \bibinfo{title}{Software testing techniques: A literature review},
\newblock in: \bibinfo{booktitle}{2016 6th international conference on
  information and communication technology for the Muslim world (ICT4M)},
  \bibinfo{organization}{IEEE}, \bibinfo{year}{2016}, pp.
  \bibinfo{pages}{177--182}.
\bibitem[{Zeller and Hildebrandt(2002)}]{Zeller02:Delta}
\bibinfo{author}{A.~Zeller}, \bibinfo{author}{R.~Hildebrandt},
\newblock \bibinfo{title}{Simplifying and isolating failure-inducing input},
\newblock \bibinfo{journal}{IEEE Transactions on Software Engineering}
  \bibinfo{volume}{28} (\bibinfo{year}{2002}) \bibinfo{pages}{183--200}.
  \DOIprefix\doi{10.1109/32.988498}.
\bibitem[{Weyuker and Jeng(1991)}]{weyuker1991analyzing}
\bibinfo{author}{E.~Weyuker}, \bibinfo{author}{B.~Jeng},
\newblock \bibinfo{title}{Analyzing partition testing strategies},
\newblock \bibinfo{journal}{IEEE Trans. on Software Engineering}
  \bibinfo{volume}{17} (\bibinfo{year}{1991}) \bibinfo{pages}{703--711}.
\bibitem[{Morell(1990)}]{morell1990theory}
\bibinfo{author}{L.~Morell},
\newblock \bibinfo{title}{A theory of fault-based testing},
\newblock \bibinfo{journal}{IEEE Transactions on Software Engineering}
  \bibinfo{volume}{16} (\bibinfo{year}{1990}) \bibinfo{pages}{844--857}.
\bibitem[{Yoo and Harman(2012)}]{yoo2012regression}
\bibinfo{author}{S.~Yoo}, \bibinfo{author}{M.~Harman},
\newblock \bibinfo{title}{Regression testing minimization, selection and
  prioritization: a survey},
\newblock \bibinfo{journal}{Software testing, verification and reliability}
  \bibinfo{volume}{22} (\bibinfo{year}{2012}) \bibinfo{pages}{67--120}.
\bibitem[{Arcuri and Briand(2011)}]{arcuri2011random}
\bibinfo{author}{A.~Arcuri}, \bibinfo{author}{L.~C. Briand},
\newblock \bibinfo{title}{Adaptive random testing: An illusion of
  effectiveness?},
\newblock in: \bibinfo{booktitle}{ISSTA}, \bibinfo{year}{2011}.
\bibitem[{Catal(2011)}]{catal2011software}
\bibinfo{author}{C.~Catal},
\newblock \bibinfo{title}{Software fault prediction: A literature review and
  current trends},
\newblock \bibinfo{journal}{Expert systems with applications}
  \bibinfo{volume}{38} (\bibinfo{year}{2011}) \bibinfo{pages}{4626--4636}.
\bibitem[{Chen et~al.(2020)Chen, Patra, Pradel, Xiong, Zhang, Hao, and
  Zhang}]{chen2020survey}
\bibinfo{author}{J.~Chen}, \bibinfo{author}{J.~Patra},
  \bibinfo{author}{M.~Pradel}, \bibinfo{author}{Y.~Xiong},
  \bibinfo{author}{H.~Zhang}, \bibinfo{author}{D.~Hao},
  \bibinfo{author}{L.~Zhang},
\newblock \bibinfo{title}{A survey of compiler testing},
\newblock \bibinfo{journal}{ACM Computing Surveys (CSUR)} \bibinfo{volume}{53}
  (\bibinfo{year}{2020}) \bibinfo{pages}{1--36}.
\bibitem[{Rumbaugh et~al.(1998)Rumbaugh, Jacobson, and Booch}]{UMLReference}
\bibinfo{author}{J.~Rumbaugh}, \bibinfo{author}{I.~Jacobson},
  \bibinfo{author}{G.~Booch}, \bibinfo{title}{The Unified Modeling Language
  Reference Manual}, Object Technology Series,
  \bibinfo{publisher}{Addison-Wesley}, \bibinfo{year}{1998}.
\bibitem[{Copeland(2004)}]{copeland2004practitioner}
\bibinfo{author}{L.~Copeland}, \bibinfo{title}{A practitioner's guide to
  software test design}, \bibinfo{publisher}{Artech House},
  \bibinfo{year}{2004}.
\bibitem[{Rothermel and Harrold(1996)}]{rothermel1996analyzing}
\bibinfo{author}{G.~Rothermel}, \bibinfo{author}{M.~J. Harrold},
\newblock \bibinfo{title}{Analyzing regression test selection techniques},
\newblock \bibinfo{journal}{IEEE Transactions on software engineering}
  \bibinfo{volume}{22} (\bibinfo{year}{1996}) \bibinfo{pages}{529--551}.
\bibitem[{Ahmed et~al.(2020)Ahmed, Enoiu, Afzal, and
  Zamli}]{ahmed2020evaluation}
\bibinfo{author}{B.~S. Ahmed}, \bibinfo{author}{E.~Enoiu},
  \bibinfo{author}{W.~Afzal}, \bibinfo{author}{K.~Z. Zamli},
\newblock \bibinfo{title}{An evaluation of monte carlo-based hyper-heuristic
  for interaction testing of industrial embedded software applications},
\newblock \bibinfo{journal}{Soft Computing} \bibinfo{volume}{24}
  (\bibinfo{year}{2020}) \bibinfo{pages}{13929--13954}.
\bibitem[{Jackson(2019)}]{jackson2019alloy}
\bibinfo{author}{D.~Jackson},
\newblock \bibinfo{title}{Alloy: a language and tool for exploring software
  designs},
\newblock \bibinfo{journal}{Communications of the ACM} \bibinfo{volume}{62}
  (\bibinfo{year}{2019}) \bibinfo{pages}{66--76}.
\bibitem[{Zeller(2001)}]{zeller2001automated}
\bibinfo{author}{A.~Zeller},
\newblock \bibinfo{title}{Automated debugging: Are we close?},
\newblock \bibinfo{journal}{Computer} \bibinfo{volume}{34}
  (\bibinfo{year}{2001}) \bibinfo{pages}{26--31}.
\bibitem[{Chen et~al.(2010)Chen, Kuo, Merkel, and Tse}]{chen2010adaptive}
\bibinfo{author}{T.~Y. Chen}, \bibinfo{author}{F.-C. Kuo},
  \bibinfo{author}{R.~G. Merkel}, \bibinfo{author}{T.~Tse},
\newblock \bibinfo{title}{Adaptive random testing: The art of test case
  diversity},
\newblock \bibinfo{journal}{Journal of Systems and Software}
  \bibinfo{volume}{83} (\bibinfo{year}{2010}) \bibinfo{pages}{60--66}.
\bibitem[{Su et~al.(2017)Su, Wu, Miao, Pu, He, Chen, and Su}]{su2017survey}
\bibinfo{author}{T.~Su}, \bibinfo{author}{K.~Wu}, \bibinfo{author}{W.~Miao},
  \bibinfo{author}{G.~Pu}, \bibinfo{author}{J.~He}, \bibinfo{author}{Y.~Chen},
  \bibinfo{author}{Z.~Su},
\newblock \bibinfo{title}{A survey on data-flow testing},
\newblock \bibinfo{journal}{ACM Computing Surveys (CSUR)} \bibinfo{volume}{50}
  (\bibinfo{year}{2017}) \bibinfo{pages}{1--35}.
\bibitem[{RTCA/DO-178C(2012)}]{DO178B}
\bibinfo{author}{RTCA/DO-178C}, \bibinfo{title}{Software considerations in
  airborne systems and equipment certification}, \bibinfo{year}{2012}.
\bibitem[{Hong et~al.(2012)Hong, Ahn, Park, Kim, and Harrold}]{hong2012testing}
\bibinfo{author}{S.~Hong}, \bibinfo{author}{J.~Ahn}, \bibinfo{author}{S.~Park},
  \bibinfo{author}{M.~Kim}, \bibinfo{author}{M.~J. Harrold},
\newblock \bibinfo{title}{Testing concurrent programs to achieve high
  synchronization coverage},
\newblock in: \bibinfo{booktitle}{Proceedings of the 2012 International
  Symposium on Software Testing and Analysis}, \bibinfo{year}{2012}, pp.
  \bibinfo{pages}{210--220}.
\bibitem[{Douglas-Smith et~al.(2020)Douglas-Smith, Iwanaga, Croke, and
  Jakeman}]{douglas2020certain}
\bibinfo{author}{D.~Douglas-Smith}, \bibinfo{author}{T.~Iwanaga},
  \bibinfo{author}{B.~F. Croke}, \bibinfo{author}{A.~J. Jakeman},
\newblock \bibinfo{title}{Certain trends in uncertainty and sensitivity
  analysis: An overview of software tools and techniques},
\newblock \bibinfo{journal}{Environmental Modelling \& Software}
  \bibinfo{volume}{124} (\bibinfo{year}{2020}) \bibinfo{pages}{104588}.
\bibitem[{Bozkurt et~al.(2010)Bozkurt, Harman, Hassoun et~al.}]{Bozkurt10:Web}
\bibinfo{author}{M.~Bozkurt}, \bibinfo{author}{M.~Harman},
  \bibinfo{author}{Y.~Hassoun}, et~al.,
\newblock \bibinfo{title}{Testing web services: A survey},
\newblock \bibinfo{journal}{Department of Computer Science, King’s College
  London, Tech. Rep. TR-10-01}  (\bibinfo{year}{2010}).
\bibitem[{Qureshi and Nadeem(2013)}]{Qureshi13:gui}
\bibinfo{author}{I.~A. Qureshi}, \bibinfo{author}{A.~Nadeem},
\newblock \bibinfo{title}{Gui testing techniques: a survey},
\newblock \bibinfo{journal}{International Journal of Future computer and
  communication} \bibinfo{volume}{2} (\bibinfo{year}{2013})
  \bibinfo{pages}{142}.
\bibitem[{Murphy-Hill et~al.(2012)Murphy-Hill, Parnin, and
  Black}]{MurphyHill12:How}
\bibinfo{author}{E.~Murphy-Hill}, \bibinfo{author}{C.~Parnin},
  \bibinfo{author}{A.~P. Black},
\newblock \bibinfo{title}{How we refactor, and how we know it},
\newblock \bibinfo{journal}{IEEE Transactions on Software Engineering}
  \bibinfo{volume}{38} (\bibinfo{year}{2012}) \bibinfo{pages}{5--18}.
  \DOIprefix\doi{10.1109/TSE.2011.41}.
\bibitem[{Martinez et~al.(2017)Martinez, Durieux, Sommerard, Xuan, and
  Monperrus}]{Martinez17:APRD4J}
\bibinfo{author}{M.~Martinez}, \bibinfo{author}{T.~Durieux},
  \bibinfo{author}{R.~Sommerard}, \bibinfo{author}{J.~Xuan},
  \bibinfo{author}{M.~Monperrus},
\newblock \bibinfo{title}{Automatic repair of real bugs in java: a large-scale
  experiment on the defects4j dataset},
\newblock \bibinfo{journal}{Empirical Software Engineering}
  \bibinfo{volume}{22} (\bibinfo{year}{2017}) \bibinfo{pages}{1936--1964}.
  \URLprefix \url{https://doi.org/10.1007/s10664-016-9470-4}.
  \DOIprefix\doi{10.1007/s10664-016-9470-4}.
\bibitem[{{Helali Moghadam} et~al.(2019){Helali Moghadam}, {Saadatmand},
  {Borg}, {Bohlin}, and {Lisper}}]{Moghadam19:RL}
\bibinfo{author}{M.~{Helali Moghadam}}, \bibinfo{author}{M.~{Saadatmand}},
  \bibinfo{author}{M.~{Borg}}, \bibinfo{author}{M.~{Bohlin}},
  \bibinfo{author}{B.~{Lisper}},
\newblock \bibinfo{title}{Machine learning to guide performance testing: An
  autonomous test framework},
\newblock in: \bibinfo{booktitle}{2019 IEEE International Conference on
  Software Testing, Verification and Validation Workshops (ICSTW)},
  \bibinfo{year}{2019}, pp. \bibinfo{pages}{164--167}.
\bibitem[{Whalen et~al.(2006)Whalen, Rajan, and
  Heimdahl}]{Whalen06:CovMetrics-ReqBased}
\bibinfo{author}{M.~Whalen}, \bibinfo{author}{A.~Rajan},
  \bibinfo{author}{M.~Heimdahl},
\newblock \bibinfo{title}{Coverage metrics for requirements-based testing},
\newblock in: \bibinfo{booktitle}{Proceedings of Int'l Symposium on Software
  Testing and Analysis}, \bibinfo{publisher}{ACM}, \bibinfo{year}{2006}, pp.
  \bibinfo{pages}{25--36}.
\bibitem[{Takanen et~al.(2008)Takanen, DeMott, and Miller}]{Takanen08:fuzzing}
\bibinfo{author}{A.~Takanen}, \bibinfo{author}{J.~DeMott},
  \bibinfo{author}{C.~Miller}, \bibinfo{title}{Fuzzing for Software Security
  Testing and Quality Assurance}, \bibinfo{publisher}{Artech House, Inc.},
  \bibinfo{year}{2008}.
\bibitem[{E.~M.~Clarke(1986)}]{Clarke86}
\bibinfo{author}{A.~S. E.~M.~Clarke, E.A.~Emerson},
\newblock \bibinfo{title}{Automatic verification of finite-state concurrent
  systems using temporal logic specifications},
\newblock \bibinfo{journal}{ACM Transactions on Programming Languages and
  Systems}  (\bibinfo{year}{1986}) \bibinfo{pages}{244--263}.
\bibitem[{Alshahwan et~al.(2018)Alshahwan, Gao, Harman, Jia, Mao, Mols, Tei,
  and Zorin}]{Alshahwan18:Sap}
\bibinfo{author}{N.~Alshahwan}, \bibinfo{author}{X.~Gao},
  \bibinfo{author}{M.~Harman}, \bibinfo{author}{Y.~Jia},
  \bibinfo{author}{K.~Mao}, \bibinfo{author}{A.~Mols},
  \bibinfo{author}{T.~Tei}, \bibinfo{author}{I.~Zorin},
\newblock \bibinfo{title}{Deploying search based software engineering with
  sapienz at facebook},
\newblock in: \bibinfo{booktitle}{Search-Based Software Engineering},
  \bibinfo{publisher}{Springer International Publishing},
  \bibinfo{address}{Cham}, \bibinfo{year}{2018}, pp. \bibinfo{pages}{3--45}.
\bibitem[{France and Rumpe(0007)}]{FoSE07:Model-Based}
\bibinfo{author}{R.~France}, \bibinfo{author}{B.~Rumpe},
\newblock \bibinfo{title}{Model-driven development of complex systems: A
  research roadmap},
\newblock in: \bibinfo{editor}{L.~Briand}, \bibinfo{editor}{A.~Wolf} (Eds.),
  \bibinfo{booktitle}{Future of Software Engineering 2007},
  \bibinfo{publisher}{IEEE-CS Press}, \bibinfo{year}{20007}.
\bibitem[{Bass et~al.(1998)Bass, Clements, and Kazman}]{Clements98:SoftArch}
\bibinfo{author}{L.~Bass}, \bibinfo{author}{P.~Clements},
  \bibinfo{author}{R.~Kazman}, \bibinfo{title}{Software Architecture in
  Practice}, \bibinfo{publisher}{Addison-Wesley}, \bibinfo{year}{1998}.
\bibitem[{Just et~al.(2011)Just, Schweiggert, and Kapfhammer}]{Just11:Major}
\bibinfo{author}{R.~Just}, \bibinfo{author}{F.~Schweiggert},
  \bibinfo{author}{G.~M. Kapfhammer},
\newblock \bibinfo{title}{Major: An efficient and extensible tool for mutation
  analysis in a java compiler},
\newblock in: \bibinfo{booktitle}{Proceedings of the 2011 26th IEEE/ACM
  International Conference on Automated Software Engineering}, ASE '11,
  \bibinfo{publisher}{IEEE Computer Society}, \bibinfo{address}{Washington, DC,
  USA}, \bibinfo{year}{2011}, pp. \bibinfo{pages}{612--615}. \URLprefix
  \url{http://dx.doi.org/10.1109/ASE.2011.6100138}.
  \DOIprefix\doi{10.1109/ASE.2011.6100138}.
\bibitem[{Fontes and Gay(2021)}]{Fontes21:oracle}
\bibinfo{author}{A.~Fontes}, \bibinfo{author}{G.~Gay},
\newblock \bibinfo{title}{Using machine learning to generate test oracles: A
  systematic literature review},
\newblock in: \bibinfo{booktitle}{Proceedings of the 1st International Workshop
  on Test Oracles}, TORACLE 2021, \bibinfo{publisher}{Association for Computing
  Machinery}, \bibinfo{address}{New York, NY, USA}, \bibinfo{year}{2021}, p.
  \bibinfo{pages}{1–10}. \URLprefix
  \url{https://doi.org/10.1145/3472675.3473974}.
  \DOIprefix\doi{10.1145/3472675.3473974}.
\bibitem[{Galeotti et~al.(2013)Galeotti, Fraser, and
  Arcuri}]{Galeotti13:search}
\bibinfo{author}{J.~P. Galeotti}, \bibinfo{author}{G.~Fraser},
  \bibinfo{author}{A.~Arcuri},
\newblock \bibinfo{title}{Improving search-based test suite generation with
  dynamic symbolic execution},
\newblock in: \bibinfo{booktitle}{2013 IEEE 24th International Symposium on
  Software Reliability Engineering (ISSRE)}, \bibinfo{year}{2013}, pp.
  \bibinfo{pages}{360--369}. \DOIprefix\doi{10.1109/ISSRE.2013.6698889}.
\bibitem[{Thelwall and Sud(2022)}]{Thelwall22:Scopus}
\bibinfo{author}{M.~Thelwall}, \bibinfo{author}{P.~Sud},
\newblock \bibinfo{title}{{Scopus 1900–2020: Growth in articles, abstracts,
  countries, fields, and journals}},
\newblock \bibinfo{journal}{Quantitative Science Studies}
  (\bibinfo{year}{2022}) \bibinfo{pages}{1--14}. \URLprefix
  \url{https://doi.org/10.1162/qss\_a\_00177}.
  \DOIprefix\doi{10.1162/qss_a_00177}.
  \href{http://arxiv.org/abs/https://direct.mit.edu/qss/article-pdf/doi/10.1162/qss\_a\_00177/1992507/qss\_a\_00177.pdf}{{\tt
  arXiv:https://direct.mit.edu/qss/article-pdf/doi/10.1162/qss\_a\_00177/1992507/qss\_a\_00177.pdf}}.
\bibitem[{Cavacini(2015)}]{Cavacini15:Scopus}
\bibinfo{author}{A.~Cavacini},
\newblock \bibinfo{title}{What is the best database for computer science
  journal articles?},
\newblock \bibinfo{journal}{Scientometrics} \bibinfo{volume}{102}
  (\bibinfo{year}{2015}) \bibinfo{pages}{2059--2071}. \URLprefix
  \url{https://doi.org/10.1007/s11192-014-1506-1}.
  \DOIprefix\doi{10.1007/s11192-014-1506-1}.
\bibitem[{Jamali et~al.(2020)Jamali, Steel, and
  Mohammadi}]{Jamali20:WineCoWord}
\bibinfo{author}{H.~Jamali}, \bibinfo{author}{C.~Steel},
  \bibinfo{author}{E.~Mohammadi},
\newblock \bibinfo{title}{Wine research and its relationship with wine
  production: a scientometric analysis of global trends},
\newblock \bibinfo{journal}{Australian Journal of Grape and Wine Research}
  \bibinfo{volume}{26} (\bibinfo{year}{2020}) \bibinfo{pages}{130--138}.
  \URLprefix \url{https://onlinelibrary.wiley.com/doi/abs/10.1111/ajgw.12422}.
  \DOIprefix\doi{https://doi.org/10.1111/ajgw.12422}.
  \href{http://arxiv.org/abs/https://onlinelibrary.wiley.com/doi/pdf/10.1111/ajgw.12422}{{\tt
  arXiv:https://onlinelibrary.wiley.com/doi/pdf/10.1111/ajgw.12422}}.
\bibitem[{Li et~al.(2016)Li, An, Wang, Huang, and Gao}]{Lo16:CoWord}
\bibinfo{author}{H.~Li}, \bibinfo{author}{H.~An}, \bibinfo{author}{Y.~Wang},
  \bibinfo{author}{J.~Huang}, \bibinfo{author}{X.~Gao},
\newblock \bibinfo{title}{Evolutionary features of academic articles co-keyword
  network and keywords co-occurrence network: Based on two-mode affiliation
  network},
\newblock \bibinfo{journal}{Physica A: Statistical Mechanics and its
  Applications} \bibinfo{volume}{450} (\bibinfo{year}{2016})
  \bibinfo{pages}{657--669}. \URLprefix
  \url{https://www.sciencedirect.com/science/article/pii/S037843711600025X}.
  \DOIprefix\doi{https://doi.org/10.1016/j.physa.2016.01.017}.
\bibitem[{Borg and Groenen(2005)}]{Borg05:Scaling}
\bibinfo{author}{I.~Borg}, \bibinfo{author}{P.~J. Groenen},
  \bibinfo{title}{Modern multidimensional scaling: Theory and applications},
  \bibinfo{publisher}{Springer Science \& Business Media},
  \bibinfo{year}{2005}.
\bibitem[{Newman(2004)}]{Newman04:Clustering}
\bibinfo{author}{M.~E. Newman},
\newblock \bibinfo{title}{Fast algorithm for detecting community structure in
  networks},
\newblock \bibinfo{journal}{Physical review E} \bibinfo{volume}{69}
  (\bibinfo{year}{2004}) \bibinfo{pages}{066133}.
\bibitem[{Waltman and van Eck(2013)}]{Waltman13:Clustering}
\bibinfo{author}{L.~Waltman}, \bibinfo{author}{N.~J. van Eck},
\newblock \bibinfo{title}{A smart local moving algorithm for large-scale
  modularity-based community detection},
\newblock \bibinfo{journal}{The European Physical Journal B}
  \bibinfo{volume}{86} (\bibinfo{year}{2013}) \bibinfo{pages}{471}. \URLprefix
  \url{https://doi.org/10.1140/epjb/e2013-40829-0}.
  \DOIprefix\doi{10.1140/epjb/e2013-40829-0}.

\end{thebibliography}

\appendix

\section{VOSViewer Technical Details}\label{sec:vosviewer}
VOSviewer produces maps based on a co-occurrence matrix---a two-dimensional matrix where each column and row represents an item---a keyword, in our case--and each cell indicates the number of times two keywords co-occur. This map construction consists of three steps. In the first step, a similarity matrix is created from the co-occurrence matrix. A map is then formed by applying the VOS mapping technique to the similarity matrix. Finally, the map is translated, rotated, and reflected.

\smallskip\noindent\textbf{Forming the similarity matrix:} 
VOSviewer takes as input a similarity matrix. This similarity matrix is obtained from the co-occurrence matrix through normalization. Normalization is done by correcting the matrix for differences in the total number of occurrences or co-occurrences of keywords. VOSviewer uses the association strength as its similarity measure~\cite{VanEck14:Visualizing}---in this case, the number of publications where two keywords are targeted together. Using the association strength, the similarity $s_{i,j}$ between two keywords $i$ and $j$ is calculated as:
\begin{equation}\label{eq:similarity}
s_{i,j} = \frac{2mc_{i,j}}{w_i w_j}
\end{equation}
where $m$ represents the total weight of all edges in the network (the total number of co-occurrences of all keywords), $c_{i,j}$ denotes the weight of the edge between keywords $i$ and $j$ (the total number of co-occurrences of the keywords), and $w_i$ and $w_j$ denote the total weight of all edges of keywords $i$ or $j$ (the total number of occurrences of keywords $i$ or $j$ and the total number of co-occurrences of these keywords with all keywords that they co-occur with). Specifically:
\begin{equation}
    w_i = \sum_{j}{c_{i,j}}
\end{equation}
\begin{equation}
    m = \frac{1}{2} \sum_{i}{w_i}
\end{equation}
\noindent The similarity between keywords $i$ and $j$ calculated using Equation~\ref{eq:similarity} is proportional to the ratio between the observed number of co-occurrences of keywords $i$ and $j$ and the expected number of co-occurrences of keywords $i$ and $j$ under the assumption that occurrences of the two keywords are independent.

\smallskip\noindent\textbf{Map formation:} The VOS mapping technique constructs a two-dimensional map in which keywords $1,...,n$ (where $n$ is the total number of keywords) are placed such that the distance between any pair of keywords $i$ and $j$ reflects their similarity $s_{i,j}$ as accurately as possible. Keywords with a high similarity are located close to each other, while keywords with a low similarity are located far from each other. 

The goal of the VOS mapping technique is to minimize the weighted sum of the squared Euclidean distances between all pairs of keywords~\cite{VanEck14:Visualizing}. The higher the similarity between the two keywords, the higher the weight of their squared distance in the summation. The specific function minimized by the mapping technique is:
\begin{equation}\label{eq:objective}
V\left(x_1, ..., x_n\right) = \sum_{i<j}{s_{i,j} \|x_i - x_j \|^2}
\end{equation}
\noindent where $x_i$ denotes the location of keyword $i$ in a two-dimensional space, and where $\|x_i - x_j\|$ denotes the Euclidean distance between keywords $i$ and $j$. To avoid trivial maps in which all keywords have the same location, minimization is subject to the constraint that the average distance between two keywords must be equal to 1. Specifically:
\begin{equation}\label{eq:constraint}
\frac{2}{n(n-1)}\sum_{i<j}{\|x_i - x_j \|=1}
\end{equation}

The constrained optimization problem---minimizing Equation~\ref{eq:objective}, subject to Equation~\ref{eq:constraint}---is solved in two steps~\cite{Van10:VOSviewer}. The constrained problem is first converted into an unconstrained problem. Second, the unconstrained problem is solved using a variant of the SMACOF algorithm, an optimization algorithm commonly used in multidimensional scaling to produce human-understandable network or graph layouts through minimization of a stress function over the positions of nodes in the graph~\cite{Borg05:Scaling}.

\smallskip\noindent\textbf{Clustering of Keywords:} Keywords are assigned to clusters, and the number of clusters is determined, through optimization. This is a common approach for clustering nodes in a network~\cite{Newman04:Clustering}. Potential assignments of keywords to clusters are assessed using the function:
\begin{equation} \label{eq:cluster}
    V(c_1,...,c_n) = \sum_{i < j}{\delta(c_i, c_j)(s_{i,j} - \gamma})
\end{equation}
\noindent where $c_i$ is the cluster that keyword $i$ has been assigned to. $\delta(c_i, c_j)$ is a difference function that yields $1$ if $c_i = c_j$ and $0$, otherwise. $\gamma$ determines the level of clustering, with higher values yielding a larger number of clusters. This equation is unified with the mapping function minimized in Equation~\ref{eq:objective}, and includes the same similarity measurement $s_{i,j}$ calculated in Equation~\ref{eq:similarity}. 

There is no maximum number of keywords per cluster. The minimum number of keywords is controlled using a user-specified parameter. We used the default, a minimum of one keyword. The clustering algorithm will merge small clusters in cases where merging does not affect the result of Equation~\ref{eq:cluster}. Therefore, any small cluster that remain are ones that affect the results of the equation. 

Equation~\ref{eq:cluster} is maximized using the smart local moving algorithm~\cite{Waltman13:Clustering}. Following the optimization, the assignment of keywords to clusters that maximizes Equation~\ref{eq:cluster} is returned. This process yields a small number of clusters containing keywords that are targeted disproportionately often together in publications.

\smallskip\noindent\textbf{Translation, rotation, and reflection:}
The optimization problem introduced in Equation~\ref{eq:objective} does not have a single global optimal solution~\cite{Van10:VOSviewer}. However, consistent results are desirable. To ensure that the same co-occurrence matrix always yields the same map, three transformations are applied after optimization:
\begin{itemize}
    \item \textbf{Translation:} The solution is translated to be centered at the origin.
    \item \textbf{Rotation:} Principle component analysis is applied in order to maximize variance on the horizontal dimension.
    \item \textbf{Reflection:} If the median of $x_{1,1},...,x_{n,1}$ is larger than 0, the solution is reflected in the vertical axis. If the median of $x_{1,2},...,x_{n,2}$ is larger than 0, the solution is reflected in the horizontal axis.
\end{itemize}

\end{document}